\documentclass[useAMS,usenatbib]{mn2e}

\usepackage{graphicx}
\usepackage{color}
\graphicspath{ {Figures/} }
\usepackage{float}
\usepackage{longtable}
\usepackage{amsmath}
\usepackage{amssymb}
\usepackage{pdflscape}
\usepackage[utf8]{inputenc}
\defcitealias{xxl_I}{XXL~paper~I}
\defcitealias{xxl_II}{XXL~paper~II}
\defcitealias{xxl_III}{XXL~paper~III}
\defcitealias{xxl_IV}{XXL~paper~IV}

\title[Dry merger driven BCG growth in XXL-100-GC]{The XXL survey XV: Evidence for dry merger driven BCG growth in XXL-100-GC X-ray clusters}
\author[S. Lavoie et al.]{\newauthor S. Lavoie$^{1, }$\thanks{E-mail: slavoie@uvic.ca}, J. P. Willis$^{1}$, J. Démoclès$^{6}$, D. Eckert$^{25}$, F. Gastaldello$^{13}$, G. P. Smith$^{2}$, \newauthor C. Lidman$^{3}$, C. Adami$^{4}$, F. Pacaud$^{5}$, M. Pierre$^{6}$, N. Clerc$^{7}$, P. Giles$^{8}$, M. Lieu$^{2}$, \newauthor L. Chiappetti$^{13}$, B. Altieri$^{9}$, F. Ardila$^{10}$, I. Baldry$^{11}$, A. Bongiorno$^{12}$, S. Desai$^{14}$,  \newauthor A. Elyiv$^{15, 20}$, L. Faccioli$^{6}$, B. Gardner$^{3, 19}$, B. Garilli$^{13}$, M. W. Groote$^{23}$, \newauthor L. Guennou$^{4, 16}$, L. Guzzo$^{17}$, A. M. Hopkins$^{3}$, J. Liske$^{22}$, S. McGee$^{2}$, O. Melnyk$^{21, 24}$, \newauthor M. S. Owers$^{3, 19}$, B. Poggianti$^{18}$, T. J. Ponman$^{2}$, M. Scodeggio$^{13}$, L. Spitler$^{3, 19}$, \newauthor R. J. Tuffs$^{23}$ \\ \\
(Affiliations can be found after the references)
}

\begin{document}
\date{Accepted 2016 August 2nd. Received ***}
\pagerange{\pageref{firstpage}--\pageref{lastpage}} \pubyear{2015}
\maketitle
\label{firstpage}

\begin{abstract}
The growth of brightest cluster galaxies is closely related to the
properties of their host cluster. We present evidence for dry mergers
as the dominant source of BCG mass growth at $z\lesssim1$ in the XXL
100 brightest cluster sample. We use the global red sequence,
H$\alpha$ emission and mean star formation history to show that
BCGs in the sample possess star formation levels comparable to field
ellipticals of similar stellar mass and redshift. XXL
100 brightest clusters are
less massive on average than those in other X-ray selected samples
such as LoCuSS or HIFLUGCS. Few clusters in the sample display high
central gas concentration, rendering inefficient the growth of BCGs
via star formation resulting from the accretion of cool gas. Using
measures of the relaxation state of their host clusters, we show
that BCGs grow as relaxation proceeds. We find that the BCG stellar
mass corresponds to a relatively constant fraction 1\%\ of the total
cluster mass in relaxed systems. We also show that, following a
cluster scale merger event, the BCG stellar mass lags behind the
expected value from the M$_{cluster}$ - M$_{BCG}$ relation but
subsequently accretes stellar mass via dry mergers as the BCG and
cluster evolve towards a relaxed state.
\end{abstract}


\section{Introduction}
Due to their dominance and location near the centre of clusters,
brightest cluster galaxy (BCG) evolution is of great interest. In the
current paradigm, BCGs are formed hierarchically by mergers with other
cluster members. Observations of $z \lesssim 0.1$ BCGs have shown that
they follow a steeper size-luminosity scaling relation than other
early-type galaxies. For their luminosity, BCGs are larger than
expected from the bulk of early-type galaxies, indicating that
dissipationless mergers play an important role in their formation
(e.g.: \citealt{bernardi_2007}; \citealt{liu_2008}). Around $z\sim1$,
BCGs gain their \emph{identity} as they unambiguously emerge as the
dominant galaxy within a cluster (\citealt{delucia_2006}). Although
early theoretical (e.g.: \citealt{merritt_84}; \citealt{merritt_85};
\citealt{schombert_87}) and more recent observational work
(\citealt{collins_2009}; \citealt{stott_2010}) favour a scenario where
BCGs were almost entirely assembled at $z\sim1$, work by
\cite{mcintosh_2008}, \cite{liu_2009} and \cite{edwards_2012} indicate
that BCGs are still growing at the present epoch. Other work by
\cite{lidman_mass}, \cite{burke_bcg} and \cite{liu_2015} indicate that
BCGs at $z<1$ still undergo major merger events and grow by a factor
of $\sim2$ from $z\sim1$ to the present epoch. Simulations have shown
that most of the mass probably comes from a small ($\lesssim10$)
number of merging events (\citealt{delucia_blaizot_2007},
\citealt{ruszkowski_2009}). Observation of mass segregation in
clusters (\citealt{dressler_97}; \citealt{adami_segregation};
\citealt{biviano_2002}; \citealt{lidman_segregation}) and the presence
of multiple bound companions around BCGs \citep{burke_bcg} show that
clusters and the BCG environment are dynamically evolving in a way
that readily makes stellar material available to BCGs.

BCG evolution is intimately linked to the host cluster evolution as
BCG growth requires an influx of material from the cluster. There are
two possible growth channels for BCGs: the accretion of stars via
gas-poor, or dry, mergers and the formation of new stars \emph{in
  situ} from accreted gas brought to the BCG by cooling flows or from
a gas-rich, or wet, merger event. Mass growth via dry mergers can only
be a major contributor to BCG mass evolution if kinematic processes in
the cluster such as dynamical friction (\citealt{chandra_friction})
make that mass available for accretion on to the BCG in timescales
less than the Hubble time. 
\cite{fabian_2012} report that most of the UV and IR luminosity of 
BCGs in cool core clusters seems to come from vigorous \emph{in situ} 
star formation, presumably fuelled by residual cooling flows.
BCG growth via such \emph{in situ} star formation requires the
host cluster to exist in a relaxed or undisturbed state as the formation of cooling
flows could be easily disrupted by cluster merging events (e.g. \citealt{ricker_2001}).

Feedback from a central active galactic nucleus can also disrupt
cooling flows via the injection of energy into the intra cluster
medium.  The duty cycle of radio-mode feedback can be more than
60\%, suppressing the amount of gas actually reaching the BCG (e.g.
\citealt{birzan_2012}). Star formation resulting from cooling flows
also requires a BCG to be situated close to the centroid of the X-ray
emission in clusters for the gas to actually be accreted
(\citealt{edwards_2007}; \citealt{bildfell_2008};
\citealt{rafferty_2008}). 

Recent work also indicate that BCGs dominant growth source changes
around $z\sim1$. \cite{webb_2015} show that very IR-luminous BCGs are
only found at $z>1$ and \cite{mcdonald_2015} find that star formation
in BCGs is more significant at $z>1$, even in dynamically disturbed
clusters. Both papers, in addition to work done by
  \cite{vulcani_2015} and \cite{liu_2013} indicate that \emph{in situ}
  star formation seems to dominate stellar mass growth at $z\gtrsim1$
  before being replaced by dry mergers at $z\lesssim1$.
Determi\textbf{\textbf{}}ning the source of BCG mass growth provides
not only a direct indication of its own evolution but also of the
history of its cluster environment.

To understand the relationship between BCGs and their host clusters
requires a large sample of such systems, ideally drawn from a range of
cluster mass and redshift, and selected according to a simple set of
physical criteria.  In this paper we investigate the properties of a
large sample of clusters and BCGs drawn from the XXL survey.  
At more than 6~Ms total exposure time over two 25
deg$^{2}$ fields, XXL is the largest \emph{XMM-Newton} programme to
date (\citealt{xxl_I}, hereafter XXL~paper~I). The two XXL survey fields are referred to as XXL-N, centred on
the XMM-LSS and CFHTLS W1 field, and XXL-S, centred on the Blanco
Cosmology Survey field. Each consists of an overlapping mosaic of 10
ks XMM exposures.

The XXL survey offers a unique perspective on the evolution of
low-to-intermediate mass X-ray clusters. Clusters
and BCGs are not homogeneous, either at fixed mass or redshift. There
are considerable variations in their properties which makes necessary
the study of a numerically large sample. The large amount of optical,
infrared and spectroscopic data available or obtained by XXL makes it
possible to study a large and well-defined X-ray cluster sample up to
$z\sim1$. More importantly, it enables us to relate photometric and
spectroscopic measures of BCGs to the relaxation state of the
clusters. We use the sample of the 100 brightest XXL clusters\footnote{Available on CDS in catalogue IX/49/xxl100gc and via the Master Catalogue Database in Milan  at: http://cosmosdb.iasf-milano.inaf.it/XXL/} for our
work (XXL-100-GC; \citealt{xxl_II}, hereafter XXL~paper~II) and find that the relaxation state of clusters
is very powerful tool to help follow and understand BCG growth.

The paper is organized as follows: in Section~\ref{sec_xxl100} we
describe the 100 brightest clusters sample and the multi-$\lambda$
data used; in Section~\ref{sec_bcg} we present the BCG selection
criteria and final sample; we present the various measurements
performed on the sample in Section~\ref{sec_measurements}; we discuss
our results in Section~\ref{sec_discussion}. A WMAP9 cosmology is used
unless otherwise stated.

\section[]{XXL-100-GC brightest clusters sample}
\label{sec_xxl100}

\subsection[]{Clusters}

Galaxy clusters are identified from processed XMM images in the
following manner: source extraction is performed by applying {\tt
  SExtractor} (\citealt{sextractor}) to wavelet-filtered XMM images.
Surface photometry is then performed on selected sources using the
custom {\sc Xamin} pipeline with sources characterized by
maximum-likelihood values of $\mathtt{extent}$,
$\mathtt{extent\_likelihood}$ and $\mathtt{detection\_likelihood}$
(\citealt{pacaud_xxl}).  The application of appropriate cuts through
this detection parameter space generate respectively the C1 cluster
sample, which is uncontaminated by misclassified sources or artefacts
(\citealt{pacaud_xxl}; \citealt{pacaud_xxl_2007}; \citealt{clerc_xxl};
\citealt{clerc_xxl_2014}) and the C2 sample which displays 30-50\%
contamination (\citealt{pierre_xxl}; \citealt{adami_xmmlss}). The
survey cluster selection function is expressed in terms of the surface
brightness of model clusters realized within XMM images
\citep{pacaud_xxl}.  A growth curve analysis is used to measure fluxes
for the 200 brightest clusters within the XXL survey footprint
(\citealt{clerc_xxl}). The analysis employs local background
estimation, nearby-source masking and interactive cluster centring.
XXL-100-GC clusters are selected from this list
with fluxes quoted in a 1$'$ radius circular aperture. The sample
contain 51 clusters located in XXL-N and 49 in XXL-S
\citepalias{xxl_II}.

Cluster X-ray temperatures for the XXL-100-GC sample are presented in
\cite{xxl_III} (hereafter XXL~paper~III). X-ray spectra of each cluster were extracted using an
aperture of radius 300~kpc with a minimum of 5 counts per spectral bin
in the 0.4-7.0~keV band. Temperatures are not core excised due to the
limited PSF of \emph{XMM-Newton} and lie mostly in the 1~KeV $\leq$
$T_{300kpc}$ $<$ 6 KeV range. 

\begin{figure}
\centering
\includegraphics[width=0.99\columnwidth]{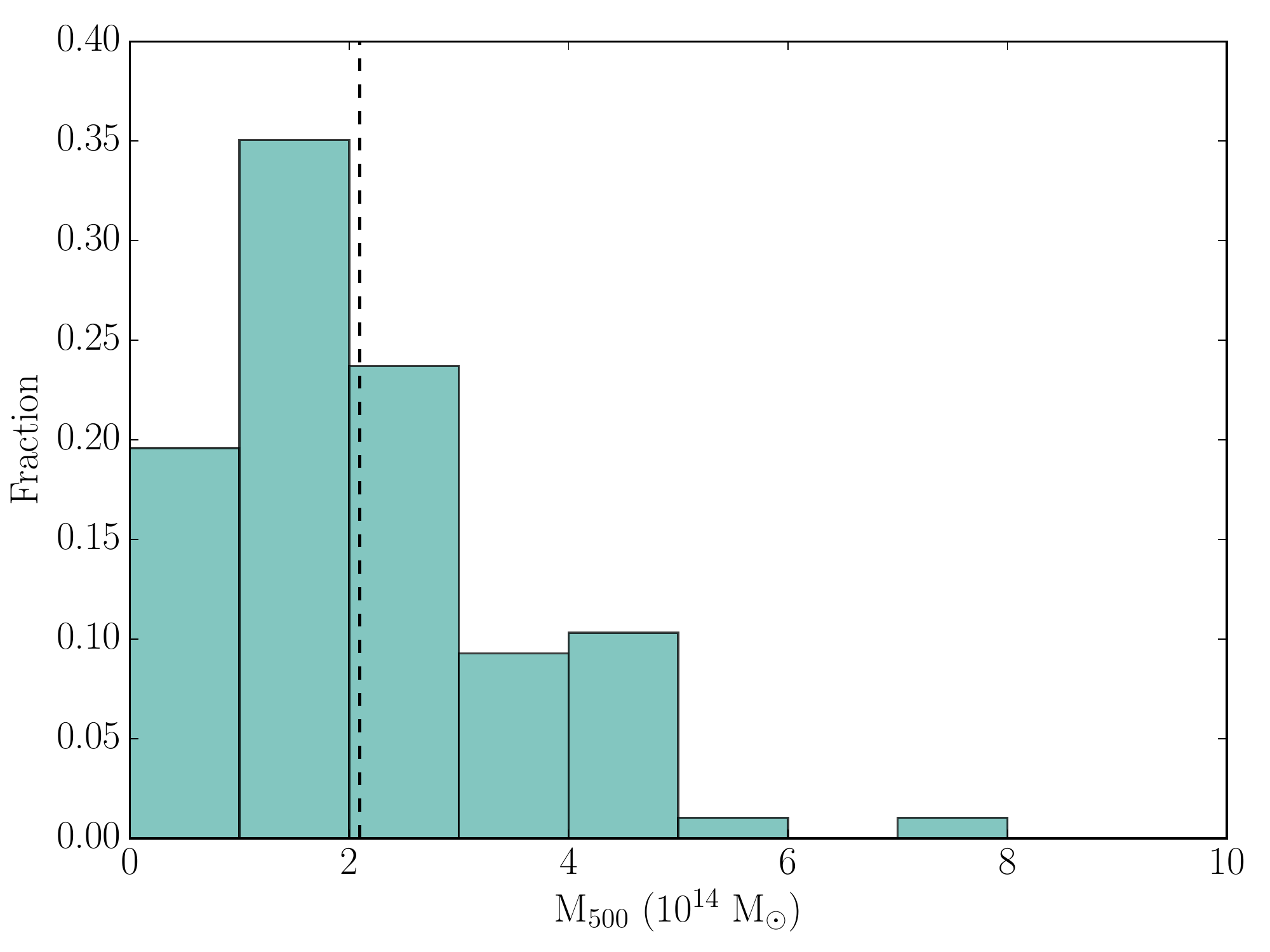}
\caption{Normalized mass distribution obtained from X-ray scaling relations
  for XXL-100-GC. The dashed vertical line indicate the average XXL-100-GC cluster mass of just over $2\times10^{14}$~M$_{\odot}$.}
\label{m_comp_fig}
\end{figure}

Cluster weak lensing masses for the XXL-100-GC sample are presented in
\cite{xxl_IV} (hereafter XXL~paper~IV). Masses are computed from an internal weak-lensing
${M-T}$ scaling relation.  calibrated using a shear profile analysis
of 38 XXL-100-GC clusters located within the footprint of the CFHTLenS
shear catalog.  Following \cite{miller_cfhtls} and
\cite{velander_cfhtls}, the authors build a shear profile from the
ellipticity analysis of galaxies found to be behind the individual
clusters in the CFHTLenS shear catalog. A Navarro, Frenk and White
(NFW; \citealt{nfw}) profile is fit to the shear profile and
integrated out to r$_{500,WL}$\footnote{Defined as the radius within
  which the average total mass density of a cluster equals 500 times
  the critical density of the Universe at the cluster redshift as
  obtained from the weak lensing analysis} to obtain the values of
weak lensing masses M$_{500,WL}$ for the clusters. The average
${M_{500,WL}-T_{300kpc}}$ scaling relation is then used to get both
r$_{500,MT}$ and M$_{500,MT}$, the mass within r$_{500}$, for all
XXL-100-GC clusters so that all masses are based on the scaling
relation. For the sake of simplicity, we shall use r$_{500}$ and
M$_{500}$ respectively to denote r$_{500,MT}$ and M$_{500,MT}$.

Figure~\ref{m_comp_fig} shows the normalized distributions of cluster
masses for XXL-100-GC as obtained from the \citetalias{xxl_IV} ${M-T}$
relation.\footnote{Most colours used in figures in this work were
  optimized for readability using the \emph{ColorBrewer} tool from
  www.ColorBrewer.org by Cynthia A. Brewer, Geography, Pennsylvania
  State University.}. The average mass within r$_{500}$ of XXL-100-GC
clusters is $\sim 2\times10^{14}$~M$_{\odot}$, a value which is
generally lower when compared to the average mass of other X-ray
cluster samples such as REXCESS ($\sim3\times10^{14}$~M$_{\odot}$,
\citealt{rexcess_paper}),
LoCuSS\footnote{http://www.sr.bham.ac.uk/locuss/home.php} ($\sim
4\times10^{14}$~M$_{\odot}$, Smith et al. 2017, in preparation), CLASH
($\sim6\times10^{14}$~M$_{\odot}$, \citealt{clash_paper}) or HIFLUGCS
($\sim 6\times10^{14}$~M$_{\odot}$, \citealt{hifglucs_2002}).
Some care must be exercised when comparing XXL-100-GC to
  samples, not just of differing mass, but also of differing sample
  selection criteria. In this sense, comparing the properties of
  XXL-100-GC to an existing, yet lower redshift, flux-limited cluster
  sample such as HIFLUGCS ($z<0.1$; \citealt{hifglucs_2002}) is of interest as it reproduces many of the selection
  biases inherent in flux- as compared to luminosity-based selection.

\subsection[]{Multiwavelength data}

XXL has been constructed as a multiwavelength survey and the complete
list of XXL-PI and external programmes can be found in
\citetalias{xxl_I}. The present work primarily employs optical and
near-infrared photometric data as well as photometric and
spectroscopic redshifts. The XXL-N field overlaps the W1 field from
CFHTLS wide MegaCam survey (\citealt{gwyn_cfhtls}). All but five of
the XXL-N clusters have \emph{ugriz} photometry from MegaCam with a
point-source \emph{i}-band depth of $\sim$25~AB. The remaining five
clusters are located in a northern extension of the CFHTLS W1 field
known as the ABC field and have \emph{grz} MegaCam photometry to the
same depth as CFHTLS.

Galaxy magnitudes are taken from the \emph{i}-band selected CFHTLS
Wide catalogue
(\citealt{gwyn_cfhtls})\footnote{http://www.cadc-ccda.hia-iha.nrc-cnrc.gc.ca/en/megapipe/cfhtls/uc.html}.
{\tt MAG\_AUTO} magnitudes in the catalogue are computed with {\tt
  SExtractor} 2.5.0 using the adaptative aperture described in
\cite{sextractor}. Extensive testing by \citeauthor{sextractor} has
shown that this aperture produces very consistent results for galaxies
of any shape or ellipticity, missing on average 6\% of the flux with
only 2\% variations rms.  We correct for the missing flux and combine
the variations with photometric errors to obtain consistent final
total magnitudes in both CFHTLS and ABC fields.
  
  W1 source photometric redshifts are taken
from the latest CFHTLS-T0007 release (\citealt{ilbert_cfhtls};
\citealt{coupon_cfhtls}) and have a typical error of
$\sigma_{W1}=0.04$ for \emph{i}$\leq$22.5.

Few sources in the ABC field have spectroscopic redshifts. Instead we
combine $grz$ photometry with the large number of sources with
spectroscopic redshifts in the W1 field to train a Generalized Linear
Models code in the ABC field (\citealt{cosmo_photoz}). The photometric
redshifts are found by passing the \emph{grz} photometry to the Python
package
\emph{CosmoPhotoz}\footnote{http://cosmophotoz.readthedocs.org}
together with the photometry and spectroscopic redshifts of about a
thousand sources in W1. This results in photometric redshifts with
$\sigma_{ABC}=0.065$ for sources with \emph{z}$\leq$23.0 in the ABC fields.

The XXL-S field is located in the sky area covered by the Blanco
Cosmology Survey (BCS) with \emph{griz} photometry
(\citealt{bcs_survey}).  Although BCS data is shallower than
  CFHTLS with a point-source \emph{i}-band depth of $~$24, the area is
  also part of the deeper Dark Energy
  Survey\footnote{http://www.darkenergysurvey.org/survey/des-description.pdf}
  (DES), a 5000~deg$^{2}$ field observed with the Dark Energy Camera
  (DECam; \citealt{decam}) in
  \emph{grizY}. While the coverage is still incomplete in the
  \emph{i}-band, it is supplemented by deeper XXL-PI observations in the \emph{grz}-band covering the Southern field with a \emph{z}-band depth of $\sim25$ (Desai et al. 2017, in preparation). DECam photometry
  is preferred whenever available. BCS magnitudes are taken from the survey
   catalogs described in \cite{bcs_sources}. Similarly, DECam data is taken from 
   the survey catalogs where total magnitudes are computed 
   from PSF corrected model fitting photometry 
  (see: \citealt{bertin_psfex} and \citealt{decam_processing}). Photometric redshifts for
sources in the Southern field are part of the BCS data and were
obtained by \cite{bcs_photoz} using the \cite{bpz} BPZ algorithm from
BCS \emph{griz} photometry. The typical photometric redshift error is
$\sigma_{BCS}=0.06$ for \emph{i}$\leq$22.5.

Spectroscopic redshifts for both XXL-N and XXL-S are drawn from a
variety of sources.  Targeted spectroscopy of individual clusters has
been obtained as part of ESO Large Programme 191.A-0268.
Further spectroscopy is available from the VIMOS
  Public Extragalacic Redshift Survey (VIPERS), a large and deep VIMOS
  (\citealt{vimos_vlt}) redshift survey focusing on the $0.5 < z <
  1.2$ redshift range (\citealt{garilli_vipers},
  \citealt{guzzo_vipers}) that partially overlaps with XXL-N.  The
  Galaxy And Mass Assembly (GAMA) survey is another large
  spectroscopic data set that overlaps XXL-N, contributing
  low-resolution, high-completeness spectroscopy of galaxies in the
  XXL-N field to $r<19.8$ (\citealt{spec_gama}, \citealt{autoz_gama},
  \citealt{dr2_gama}). Data exchange with the VIPERS and GAMA teams
has made available thousands of spectroscopic redshifts for this
work. In addition, publicly available spectroscopic redshifts from
SDSS DR12 (\citealt{sdss_gunn}, \citealt{sdss_iii},
\citealt{sdss_boss}, \citealt{sdss_mos}, \citealt{ahn_sdss}) and the
VIMOS VLT deep survey (\citealt{vimos_vlt_deep}) are used
where they overlap with XXL-N. Many smaller XXL
programmes were undertaken to complement the spectroscopic redshifts
in the Northern and Southern fields by focusing on known XXL
clusters. Most of the spectra in the South have been obtained with the
AAOmega spectrograph (\citealt{aaomega_saunders};
\citealt{aaomega_smith}) on the Anglo-Australian Telescope \citep{xxl_xiv}.  
Table~\ref{spec_sources} lists basic
information on the various sources of spectroscopic data.

\begin{table*}
\caption{Summary of spectroscopic data covering XXL-100-GC fields used for this work.}
\begin{center}
\begin{tabular}{lccccc}
\hline
Instrument/Programme & Field & Resolution & Coverage & Typical t$_{exp}$ \\
\hline
VIMOS/VIPERS  & N & R=1200 & 16 deg$^{2}$ & 2700s  \\
VIMOS/VLT deep survey & N & R=230 & 0.61 deg$^{2}$ & 16~200s \\
AAOmega/GAMA & N & R=1400 & 23.5 deg$^{2}$ overlap with XXL & 3000-5000s  \\ 
BOSS/SDSS DR12 & N & R=1300-3000 & All XXL-N & 2700s \\
AF2/XXL-PI & N & R=1200 & Individual clusters & 7200s / 14~400s \\
EFOSC2/XXL-PI & N+S & R=300 & Individual clusters & 2700s \\
FORS2/XXL-PI & N+S & R=600 & Individual clusters & 2400s \\
AAOmega/XXL-PI & S & R=1400 & 25 deg$^{2}$ & 5000-10~000s \\
\hline
\end{tabular}
\end{center}
\label{spec_sources}
\end{table*}

Since we have access to such a large number of photometric and
spectroscopic redshifts in both the North and South field, it is
possible to evaluate and correct the \emph{redshift bias}. Due to the
inherent difficulty of associating the right template to a galaxy,
photometric redshifts can show systematic offset from their
spectroscopic counterpart. One has to correct for this effect to
reliably associate galaxies with their host cluster. In XXL-100-GC data,
this effect is larger at $z_{spec}\lesssim0.1$ and
$z_{spec}\gtrsim0.8$.  Assuming that spectroscopic redshifts are
right, we build a redshift bias correction curve for each field from
the sources that have both a spectroscopic and photometric
redshift. We then apply the correction to all sources that only have a
photometric redshift and use those corrected values for this work.

\section[]{Brightest cluster galaxies}
\label{sec_bcg}

Given the availability of good quality multi-band photometry together
with photometric and spectroscopic redshifts to $z<1$, a simple set of
criteria can be used to identify BCGs. For the present work, we define
a BCG as:

\begin{itemize}
\renewcommand{\labelitemi}{$-$}
\item The brightest galaxy in \emph{z}-band,
\item within 0.5 $\times$ r$_{500}$ of the cluster X-ray centroid,
\item with a redshift that is consistent with that of the cluster as determined from all the redshifts available around the X-ray centroid.
\end{itemize}
A coarse selection of possible cluster members is first done using
photometric redshifts. Galaxies within 0.5~$\times$~r$_{500}$ of a
cluster X-ray centroid are considered possible members if their
photometric redshift falls within:
\begin{equation*}
|z_{gal} - z_{cl}| \leq \sigma_{x}\times (1 + z_{cl}),
\end{equation*}
where $z_{gal}$ is the galaxy photometric redshift, $z_{cl}$ is the
cluster redshift and $\sigma_{x}$ is the 1-$\sigma$ error on the
photometric redshift from the method used in the different fields
($\sigma_{W1}=0.04$ for \emph{i}$\leq$22.5, $\sigma_{BCS}=0.06$ for
\emph{i}$\leq$22.5 and $\sigma_{ABC}=0.065$ for
\emph{z}$\leq$23.0). The brightest \emph{z}-band galaxy from that
selection is used as a candidate BCG. In $\sim$90\% of these cases,
visual inspection confirms that the selected BCG is a sensible
choice. For the remaining $\sim$10\% of systems, photometric redshifts
are ignored and the BCG candidates are identified from photometry
alone before being visually confirmed.  Spectroscopic redshifts are
available for all but 3 BCGs and of those with spectra all are
confirmed to be $<$3000~km~s$^{-1}$ from their cluster
redshift. Additionally, all of the BCGs identified from photometry
alone have a spectroscopic redshift consistent with the host cluster.

Some XXL-100-GC clusters are excluded from this study for various
reasons. XLSSC~088, XLSSC~092, XLSSC~110, XLSSC~501, XLSSC~526 and XLSSC~536 are
excluded because the photometry of their identified BCG is possibly
contaminated by obvious foreground objects along 
the line of sight. XLSSC~089, XLSSC~094 and XLSSC~102 are excluded due 
to the lack of redshifts available to confirm selections that are dubious. 
Two additional clusters, XLSSC~504 and XLSSC~508, are excluded due to possible
contamination of their X-ray centroid from an AGN. XLSSC~052
and XLSSC~062 are excluded because they have only been observed by
CFH12K in a few bands. Additional clusters are excluded because measurements of 
their mass or X-ray relaxation are unavailable. Our final sample consists of 85 clusters, 45 of
which are in the Northern field and 40 in the Southern field. For the
sake of simplicity, XXL-100-GC will refer to those 85 clusters for the
remainder of the paper. BCG positions and some of their
characteristics determined later in the paper are presented in
Table~\ref{bcg_position}.

\section[]{Measurements}
\label{sec_measurements}
A range of measurements can be performed upon the sample to search for
evidence of a particular source of BCG growth.  We describe these in
detail in the following sections and summarize them here. The position
of the BCG in relation to the X-ray centroid of their host cluster is
measured and an estimate of the X-ray emitting gas concentration is
taken from Démoclès et al. (2017, in preparation). Both measures are employed as
indicators of the relaxation state of the clusters. The quality of the
photometry in the W1 field and the size of the BCG sample enables us
to determine the average star formation history for the BCGs. From
this model, we compute stellar masses for all XXL-100-GC BCGs. We use
photometric and spectroscopic redshifts to identify individual members
of a given cluster and measure the difference in magnitude between the
BCG and bright cluster members as well as investigate evidence of
luminosity segregation. We employ the results from a semi-analytic
simulation of galaxy evolution to obtain an insight into the
distribution of galaxy masses accreted by the BCG. Where available,
H$\alpha$ emission line fluxes are measured from SDSS DR12
spectroscopy and are employed to determine the level of ongoing star
formation in BCGs. Finally, a global red-sequence for the XXL-100-GC
cluster sample is constructed by applying appropriate $k$- and
distance modulus corrections to transform individual cluster member
photometry to a common redshift. The distance of individual BCGs from
the global red sequence is then employed to investigate the extent to
which the star formation history of individual BCGs differs from the
average properties of the cluster sample.

\subsection{BCG offset from X-ray centroid}
\label{sec_bcg_offset}

As the most massive galaxy within a cluster, the BCG migrates to the
centre of the host cluster as a result of dynamical friction.  As the
X-ray emitting gas provides an effective observational tracer of the
cluster potential, the offset between the X-ray centroid and a BCG can
be used as an indicator of the relaxation state of a cluster. In a
relaxed cluster the offset between the BCG position and the X-ray
centroid should approach zero.

\begin{figure}
\centering
\includegraphics[width=0.99\columnwidth]{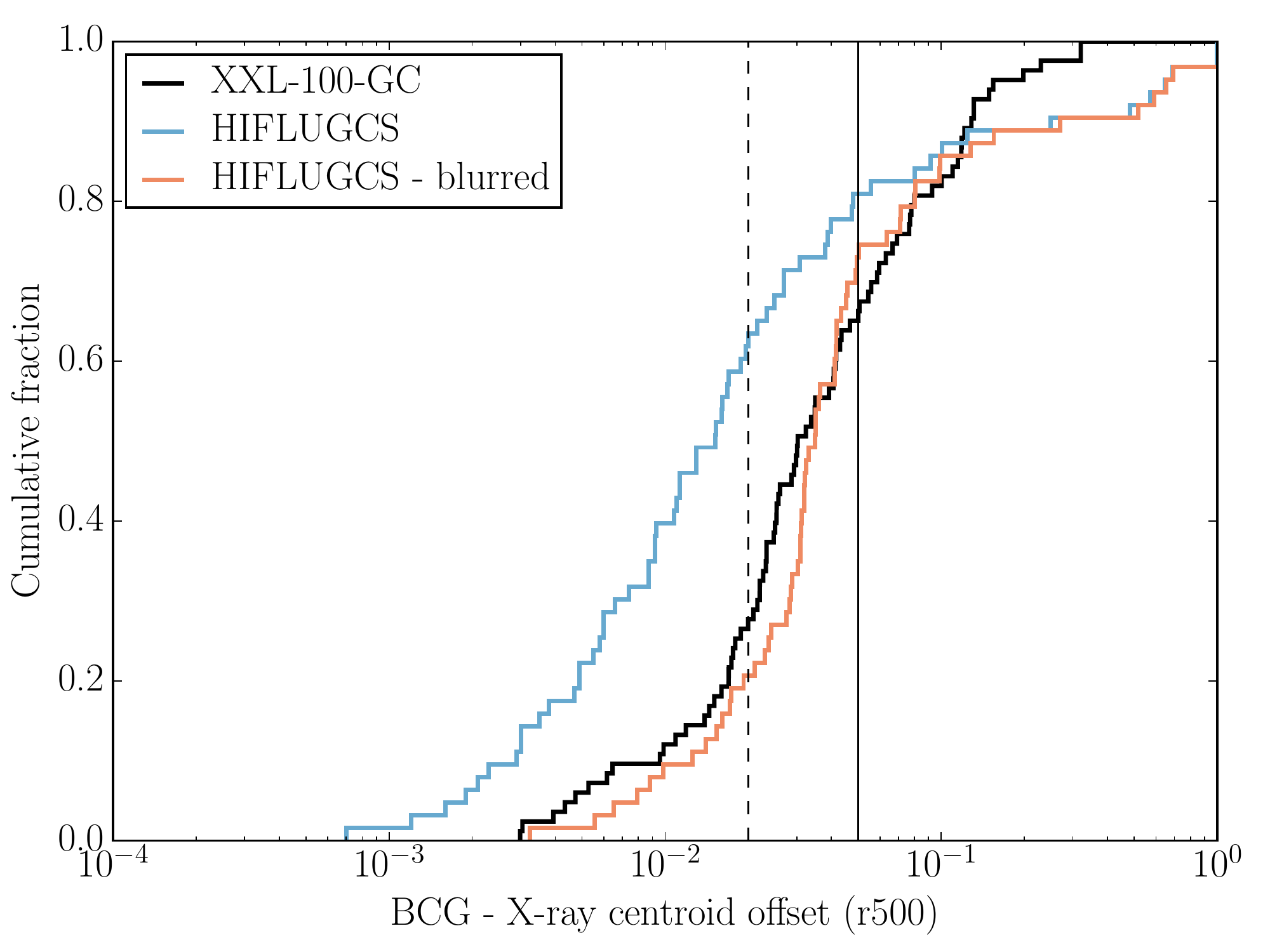}
\caption{Comparison of BCG offsets from the X-ray
    centroid for XXL-100-GC (black), HIFLUGCS (blue). The red line
    represent the HIFLUGCS offset distribution transposed to the
    median redshift of the XXL-100-GC sample ($z=0.33$) and modified 
    by a Rayleigh distribution with a scale parameter of $5"$
    applied to the X-ray centroid position. The dashed and solid
    vertical lines represent BCG offsets of $0.02\times r_{500}$ and
    $0.05\times r_{500}$, respectively.}
\label{offset_comp}
\end{figure}

We combine X-ray centroid positions and $r_{500}$
  values from \citetalias{xxl_II} with our BCG positions, to compute the
  centroid offset for the XXL-100-GC BCG sample in units of $r_{500}$
  (listed in Table~\ref{bcg_position}). Scaling the offsets by
  $r_{500}$ offers a suitable normalization method based on the mass
  distribution in each cluster. The extent of the \emph{XMM} PSF
  results in an error of approximately
  $3.6"$ (1-$\sigma$) respectively in RA and DEC in the measured X-ray centroid 
  of moderately bright ($>300$ counts), extended sources
  (Faccioli et al. 2017, in preparation). Figure~\ref{offset_comp} illustrates the
  effect of this positional error in a comparison of the distribution
  of BCG offsets in the XXL-100-GC and HIFLUGCS \citep{zhang_hiflugcs} surveys.  
  Although it appears that the XXL-100-GC sample is
  lacking in low-offset BCGs compared to HIFLUGCS, we demonstrate that
  this difference is largely a result of the centroid uncertainty of
  XXL-100-GC clusters. Figure~\ref{offset_comp} displays the HIFLUGCS
  offset distribution transposed to the median redshift ($z=0.33$) of
  the XXL-100-GC sample and modified by a Rayleigh distribution with
   a scale parameter of $5"$ (the quadratic combination of the error in both axis) applied to the X-ray centroid (red
  line).  One can see that the effect of this is to scatter low-offset
  BCGs to higher offsets, bringing the distribution into closer
  agreement with the XXL-100-GC distribution.

Despite this position error, BCG offsets may still be employed to
classify clusters as relaxed or unrelaxed.  We select a threshold of
$0.05 \times r_{500}$ as it is large enough to be unaffected by the
X-ray centroid error over the full range of XXL-100-GC redshifts yet
provides physically sensible results when applied in later analyses in
this paper.  In particular, the angular scale defined by $0.05 \times
r_{500}$ for an example cluster at $z=1$ with $r_{500}=700$~kpc is two
times larger than the angular error in the X-ray centroid.  Clusters
with a normalized BCG offset from the X-ray centroid lower than
$0.05~r_{500}$ will be considered \emph{relaxed}, while those with a
larger offset will be considered \emph{unrelaxed}. We note that we
have experimented with varying this threshold, in particular setting
the threshold for an unrelaxed cluster as $>0.1 \times r_{500}$. This
selection did not change the qualitative nature of the results
presented in this paper and resulted in a much smaller sample of
clusters classified as unrelaxed (16 instead of 30).  Physically, the
important distinction therefore appears to be to separate clusters
into low-offset, relaxed clusters and the rest.

\subsection[]{X-ray gas concentration}
\label{sec_xray_state}

Clusters that display very peaked central X-ray surface brightness
profiles may be classified as \emph{cool-core} clusters. Such cool
cores in massive clusters are associated observationally with central
concentrations of cool X-ray gas and optical emission line filaments
which appear to be accreting onto the BCG (e.g. \citealt{crawford_1999}).  
Cool core clusters can be disrupted
by cluster scale merging events as a result of the input of kinetic
energy from the merger into the cluster ICM
(e.g. \citealt{ricker_2001}).  Energy input to the central gas
concentration raises it to a higher energy state within the cluster
potential, i.e. moves it to larger clustercentric radius. In addition
to cluster merging, an AGN outburst in the BCG could also disrupt the
properties of a cool-core (e.g. \citealt{guo_2009}).  Although
observed X-ray surface brightness profiles of clusters display
considerable variation, they remain an effective indicator of the
presence of cool core within a cluster.

For clusters observed at sufficiently high resolution, the central
slope of the X-ray surface brightness profile can be used to estimate
the gas concentration and the relaxation state.  XXL-100-GC spatial
resolution is limited by the relatively large PSF of
\emph{XMM-Newton}, making the measurement of the inner slope
impractical for the whole sample. Instead we obtain the X-ray gas
concentration measurements for XXL-100-GC from Démoclès et al. (2017, in
preparation) who compute the $c_{SB}$ parameter defined by
\cite{santos_csb} as the ratio of the average surface brightness
within 40 and 400~kpc. \cite{santos_2010} and \cite{hudson_2010} show
that $c_{SB}$ has a low scatter with cluster central cooling time,
making it a reliable indicator of cluster relaxation.

\begin{figure}
\centering
\includegraphics[width=0.99\columnwidth]{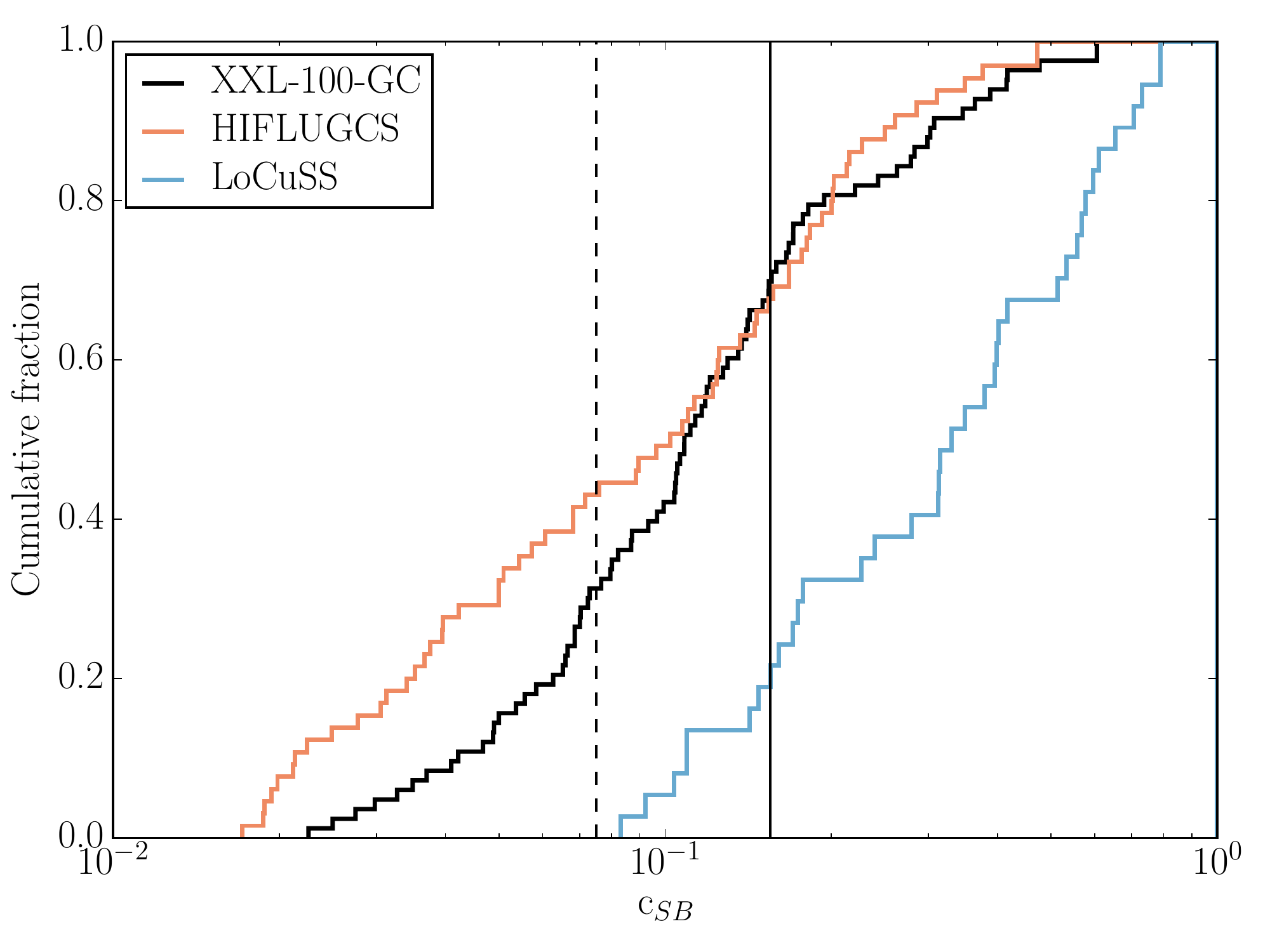}
\caption{Cumulative fraction of X-ray gas
    concentration parameter for the XXL-100-GC and HIFLUGCS samples.  The
    dashed and solid vertical lines indicate $c_{SB}=0.075$ and
    $0.155$, respectively separating each distribution into non-,
    weak- and strong-cool-core clusters according to
    \protect\cite{santos_csb}.}
\label{csb_comp}
\end{figure}

We test the robustness of the method used to measure
  the $c_{SB}$ parameter in XXL-100-GC by applying the same procedure to
  mock X-ray images created from the cosmoOWLS simulation
  \citep{lebrun_2014}. CosmoOWLS is a large suite of smoothed particle
  hydrodynamics (SPH) simulations within a cosmological volume that
  include the effects of a variety of gas physics, such as gas cooling
  and feedback from supernovae and active galactic nuclei (AGN).
  Simulated X-ray images of 25 clusters spanning the whole range of
  $c_{SB}$ with similar redshift and temperature distributions to
  XXL-100-GC were created. The simulated images were then folded through
  the \emph{XMM-Newton} response and convolved with the PSF of the
  telescope. A realistic background was added to the images to create
  a mock \emph{XMM-Newton} image similar to real XXL
  observations. Finally, the same method was applied to measure the
  concentration parameter of both the mock images and the original
  simulated data.  The median value of $c_{\rm SB, mock} - c_{\rm SB,
    true}$ is $-0.02$ with a scatter of 0.13. Therefore, our method is
  able to recover the concentration of XXL clusters in a relatively
  unbiased manner albeit with limited precision. As a final check, in
  Figure~\ref{csb_comp} we compare the distribution of $c_{SB}$
  values measured for the XXL-100-GC sample to the HIFLUGCS sample
  \citep{hudson_2010}. Both samples are area-complete and
  flux-limited, yet with different mean redshifts, and display
  $c_{SB}$ distributions that are qualitatively very similar. To highlight how 
  different selection methods affect the resulting sample, 
  an estimate (without PSF correction) of the $c_{SB}$ values for the luminosity-selected 
  LoCuSS sample is also shown (Démoclès, Smith \& Martino, private communications). 

\subsection[]{BCG stellar masses}
\label{sec_bcg_mass}
Estimating the stellar mass of a BCG requires knowledge of its star
formation history. Following \cite{lidman_mass}, we deduce an average
star formation history (SFH) for the BCG sample. We then employ
this global SFH to estimate the stellar masses of individual BCGs. As
the MegaCam photometry in the W1 field is the most reliable, we derive
the SFH of the sample using only these BCGs. This SFH model is then
applied to the whole sample assuming that the BCGs in the BCS and
DECam fields are physically identical on average to the ones in
W1. Extinction in the W1 field is low ($\sim0.03$ \emph{z}-mag and
$\sim0.01$~\emph{i-z} colour) and is ignored as model uncertainties
dominate. W1 photometry has a typical night to night scatter of 0.03
mag that is combined quadratically with each BCGs photometric
uncertainty.

\begin{figure}
\centering
\includegraphics[width=0.99\columnwidth]{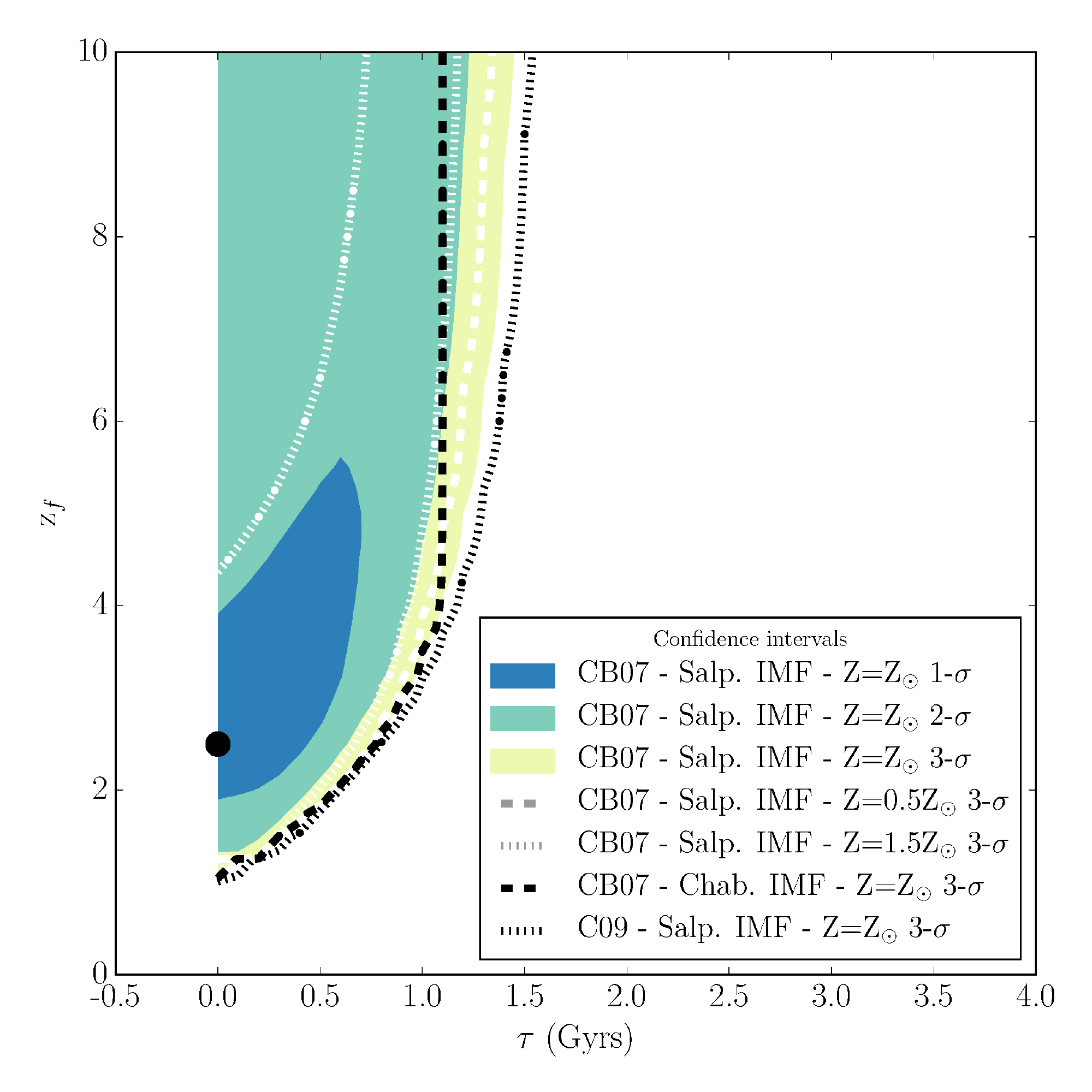}
\caption{Star formation history confidence intervals contours. The 1-,
  2- and 3-$\sigma$ contours (respectively shown in blue, green and
  yellow) representing the quality of the fit between observed BCGs
  \emph{i-z} colour evolution and a CB07 model with Z=Z$_{\odot}$ and
  Salpeter IMF. The other dashed lines represent various 3-$\sigma$
  contours obtained for different IMF or metallicity choices. The fit
  with the lowest $\chi^{2}$ value is represented by the black dot and
  corresponds to CB07 model with a single-stellar population, solar
  metallicity and Salpeter IMF.}
\label{SFH-contours}
\end{figure}

\begin{figure*}
\centering
\includegraphics[width=0.99\textwidth]{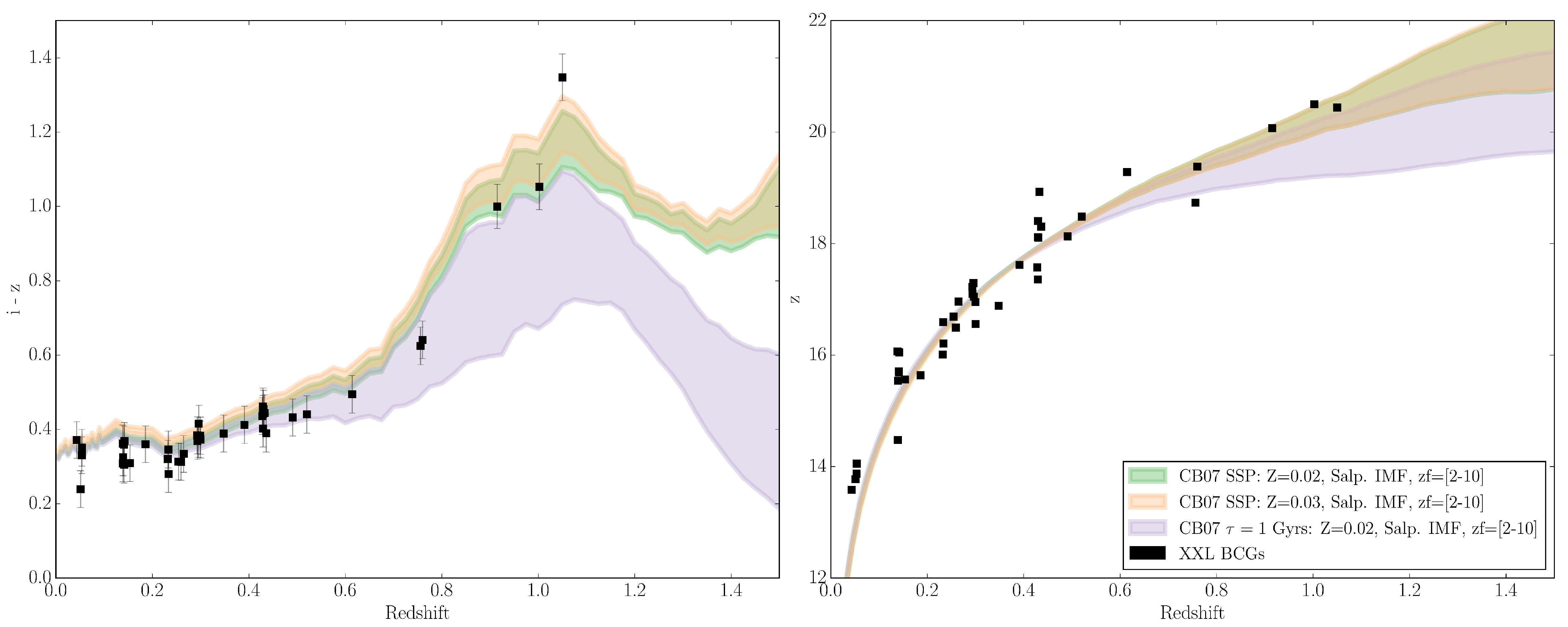}
\caption{Colour (\emph{left}) and \emph{z}-band magnitude
  (\emph{right}) evolution of the XXL-100-GC North subsample. The green
  band represent the evolution of the best fitting model for
  \emph{z$_{f}$} between 2 and 10 with solar metallicity and Salpeter
  IMF. The red band is the same model with $1.5\times$~Z$_{\odot}$ and
  the blue band is a model with solar metallicity and
  $\tau=1$~Gyrs. The black squares represent XXL-100-GC North BCGs. Both
  panels use the same legend.}
\label{SFH-color}
\end{figure*}

We determine the best-fitting SFH model for the XXL-100-GC sample by
comparing model stellar populations of varying properties to the
sample of BCG colours versus redshift.  We employ the
\emph{EzGal}\footnote{http://www.baryons.org/ezgal/} Python package to
produce the stellar population models and determine the model that
best reproduces the observed \emph{i-z} colours of the W1 BCGs
sub-sample.  This is achieved by identifying the lowest weighted
$\chi^{2}$ value for a set of models with metallicity Z=0.5, 0.75, 1,
1.5 and 2.5 Z$_{\odot}$. For each metallicity value, the best fitting
model is sought by varying the timescale $\tau$ of an $e$-folding
model star formation rate between 0 and 10~Gyrs and the formation
redshift (z$_{f}$) between 1 and 5.  The process is performed for both
a Charlot-Bruzual (CB07) model (\citealt{bc03}; Charlot \& Bruzual, in
preparation) with a Salpeter initial mass function (IMF; \citealt{salpeter})
and a Chabrier IMF (\citealt{chabrier}). A family of Conroy (C09)
models (\citealt{c09}) with Z=Z$_{\odot}$ and a Salpeter IMF is also
used to see if the choice of model greatly influences our final SFH.

Figure~\ref{SFH-contours} displays the confidence intervals obtained
for each set of models. The various models differ little in the
parameter space enclosed by their 3-$\sigma$ confidence interval and
minimal $\chi^{2}$ values.  We select a CB07 model described by a
Salpeter IMF with Z=Z$_{\odot}$, single stellar population (SSP) and
$z_{f}$=2.5 as our average star formation history because it has the
lowest formal $\chi^{2}$ and this metallicity has a slightly more
precisely defined 1-$\sigma$ confidence interval.  We prefer the use
of a Salpeter IMF as \cite{imf_selection} demonstrate that a
``bottom-heavy'' IMF potentially provides a better description of the
SFH of massive galaxies than a ``bottom-light'' Chabrier IMF. Our
findings are slightly different than those reported by
\cite{lidman_mass} yet overall agree at the 2-$\sigma$ confidence
level.  Furthermore \cite{lidman_mass} also employ \emph{J-K} colours
to constrain the average star formation history, which is less
sensitive to recent star formation than our \emph{i-z} colours.

Figure~\ref{SFH-color} indicates how metallicity, star formation
history and the redshift of formation affect the predicted values of
colour and \emph{z}-band magnitude. Maintaining the same star
formation history, one notes that though metallicity variations act to
offset the predicted colours they do not significantly alter the
\emph{z}-band magnitude, our proxy for stellar mass. A star formation
history with a non-zero $\tau$ generates bluer galaxy colours at high
redshift. Even a small positive value of $\tau$ is in tension with the
colours observed for high-\emph{z} BCGs in XXL-100-GC, pointing towards
passive evolution since early times. We note that none of the models works 
completely, some BCGs being bluer in \emph{i-z} than any of our model can 
reproduce. 

Employing the best fitting model within our adopted cosmology, we
apply a simple bisection algorithm to obtain the absolute
\emph{z}-band magnitude that best reproduces the observed magnitude of
each BCG. DECam \emph{z}-band magnitudes are used for BCGs in
  the XXL-S field. Stellar masses are obtained by applying a
mass-to-light ratio appropriate for the SFH model to each BCG $z$-band
luminosity.  The effect of switching between an assumed Chabrier or
Kroupa IMF is to change the derived stellar masses equally over the
XXL-100-GC sample without introducing any IMF-dependent evolution with
redshift.  The influence on the relations derived from BCG masses is
also marginal. The uncertainties in Table~\ref{bcg_position} represent
the range of masses within the 1-$\sigma$ confidence interval shown in
Figure~\ref{SFH-contours}. As one can see from the shaded
  regions in Figure~\ref{SFH-color}, model errors become more
  important at higher redshift. Because of this, model errors dominate
  mass uncertainties for all BCGs. Resulting mass uncertainties are
  $\sim10-20\%$, somewhat higher than the $\sim5-10\%$ obtained with a
  similar method by \cite{lidman_mass} without model errors.
Additionally, masses obtained from a reprocessing by the Portsmouth
Group of SDSS
DR12\footnote{http://www.sdss.org/dr12/spectro/galaxy\_portsmouth/}
following \cite{maraston_sdss_masses} are available for 30
BCGs. Values determined for our BCGs fall within $\pm$0.2~dex of the
masses they report for a passive model with Salpeter IMF.
 
\subsection[]{M$_{cluster}$ - M$_{BCG}$ relation}

It is generally accepted that more massive BCGs exist in more massive
clusters. In the hierarchical scenario, as massive clusters grow by
the accretion of less massive sub-units, recently accreted galaxies
migrate to the centre of the cluster potential where they are
themselves accreted by the BCG which itself grows in mass.  The
relationship between cluster and BCG mass may be expressed as a simple
power law relationship of the form $M_{cluster}=A~\times M_{BCG}^{n}$.
Various measurements of the power law exponent $n$ for can be found in
the literature for cluster samples typically limited in mass to
$M_{cluster}>10^{14}$~M$_{\odot}$.  \cite{stott_2010} find a power law
index of $2.4\pm0.6$ for a sample of 20 $z>0.8$ X-ray luminous
clusters identified from either their X-ray emission or various
optical methods. \cite{stott_2012} obtain an index of $1.3\pm0.1$ from
103 clusters chosen from the XCS first release by applying a redshift
cutoff of $z<0.3$. Finally, \cite{lidman_mass} combine data from
\cite{stott_2010} and \cite{stott_2012} with a sample of SpARCS
clusters identified as galaxy overdensities in deep IR observations to
obtain an index of $1.6\pm0.2$.

XXL-100-GC provides an important perspective on the relationship between
BCG and cluster masses as it samples a range of clusters masses
typically lower than those studied in the literature and because it
includes additional diagnostic information on the relaxation state of
each cluster.  For the purpose of this analysis we assume that a
relaxed cluster is either characterised by the presence of a
cool-core, indicated by a high value of c$_{SB}$, or a dynamically
relaxed BCG, indicated by a low value of normalised offset from the
centroid.

\begin{figure}
\centering
\includegraphics[width=0.99\columnwidth]{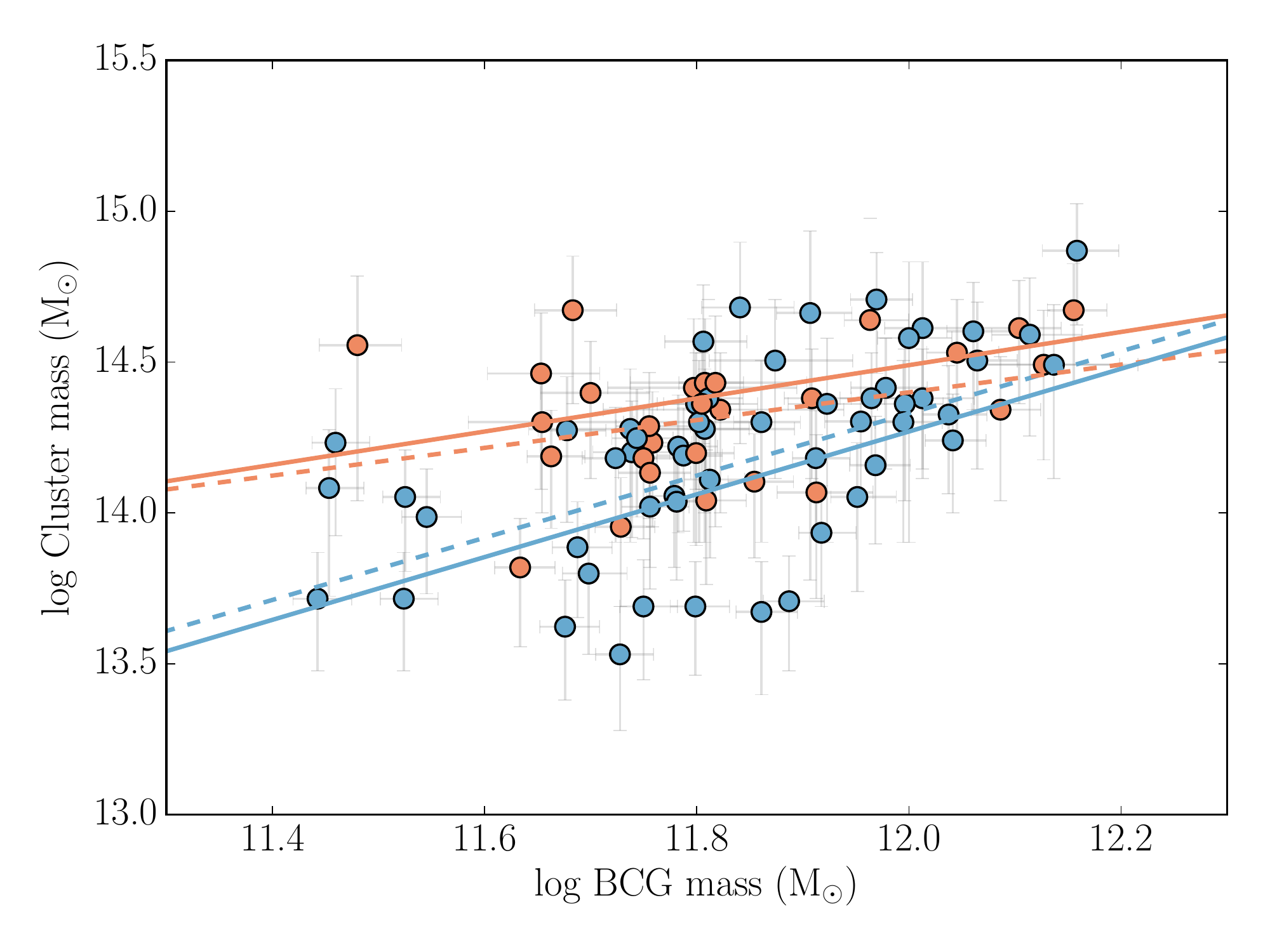}
\caption{Cluster mass versus BCG mass for XXL-100-GC clusters. Points are
  colour-coded to emphasize the distribution of clusters exhibiting
  different relaxation states. The two colours for dots represent
  BCG offsets of $<0.05\times r_{500}$ (blue) and $>0.05\times r_{500}$
  (red).  The dashed blue and red lines
  respectively indicate the result of linear fits to clusters of BCG
  offset $<0.05\times r_{500}$ and $>0.05\times r_{500}$.  The solid
  blue and red lines respectively indicate the result of linear fits
  to clusters of weighted by the value of c$_{SB}$ and inverse
  c$_{SB}$.}
\label{cl_bcg-mass}
\end{figure}

We therefore perform a number of fits to the slope of $M_{BCG}$ versus
$M_{cluster}$ employing different assumptions. The best fit was
obtained employing $\chi^{2}$ minimization and resampling the data
100,000 times assuming data uncertainties are normally distributed,
taking the median index value and standard deviation. The results are
indicated in Table \ref{clm_bcm_table} and in Figure
\ref{cl_bcg-mass}. We perform an unweighted fit to the mass data
points to provide a baseline description of the relationship.  We also
perform a fit employing cluster c$_{SB}$ values as a simple weighting
function in order to weight the contribution of relaxed clusters in
the relationship.  A similar fit employing inverse c$_{SB}$ values
weights the relationship toward unrelaxed clusters.  Finally we also
perform fits using only clusters with BCGs located a small or large
offset radii to perform an alternative description of the relationship
for relaxed or unrelaxed clusters respectively.

The fit results for BCGs located in relaxed clusters is consistent
with the scenario where the BCG stellar mass is proportional to the
total cluster mass.  The fit normalisation is such that the BCG
stellar mass represents an approximately constant 1\%\ of the total
cluster mass.  The fit results also indicate that, at fixed cluster
mass, BCGs in unrelaxed clusters are less massive than BCGs in relaxed
clusters by up to 0.5~dex.  This impression is characterized by the
trend for BCGs in unrelaxed clusters to lie predominantly to the left
of the $M_{BCG}$ versus $M_{cluster}$ relationship defined by relaxed
clusters as shown in Figure \ref{cl_bcg-mass}.  To test the
significance of this trend, we define a simple normalized distance
measure from the relaxed relation. For each BCG, we measure the
distance between the expected BCG mass at the host cluster mass,
normalized by the BCG mass (denoted $\Delta M/M_{BCG}$). In other
words, the difference between the BCG mass and how massive is the BCG
expected to be if it were in a relaxed cluster with its host cluster
mass.

\begin{table}
\begin{center}
\caption{Properties of the $M_{cluster}=A~\times M_{BCG}^{n}$ fits
  shown in Figure~\ref{cl_bcg-mass}.}
\label{clm_bcm_table}	
\begin{tabular}{lccc}
\hline
Case & $n$ & log$A$  & \#\\
\hline
All clusters & 0.84$\pm0.09$ & 4.33 & 85\\
c$_{SB}$ weighted  & 1.04$\pm0.24$ & 1.79 & -\\
(c$_{SB}$)$^{-1}$ weighted & 0.55$\pm0.16$ & 7.89 & -\\
BCG offset $<0.05\times r_{500}$ & 1.03$\pm0.10$ & 1.97 & 55\\
BCG offset $>0.05\times r_{500}$ & 0.46$\pm0.17$ & 8.88 & 30\\
\hline
\end{tabular}
\end{center}
\end{table}

\begin{figure}
\centering
\includegraphics[width=0.99\columnwidth]{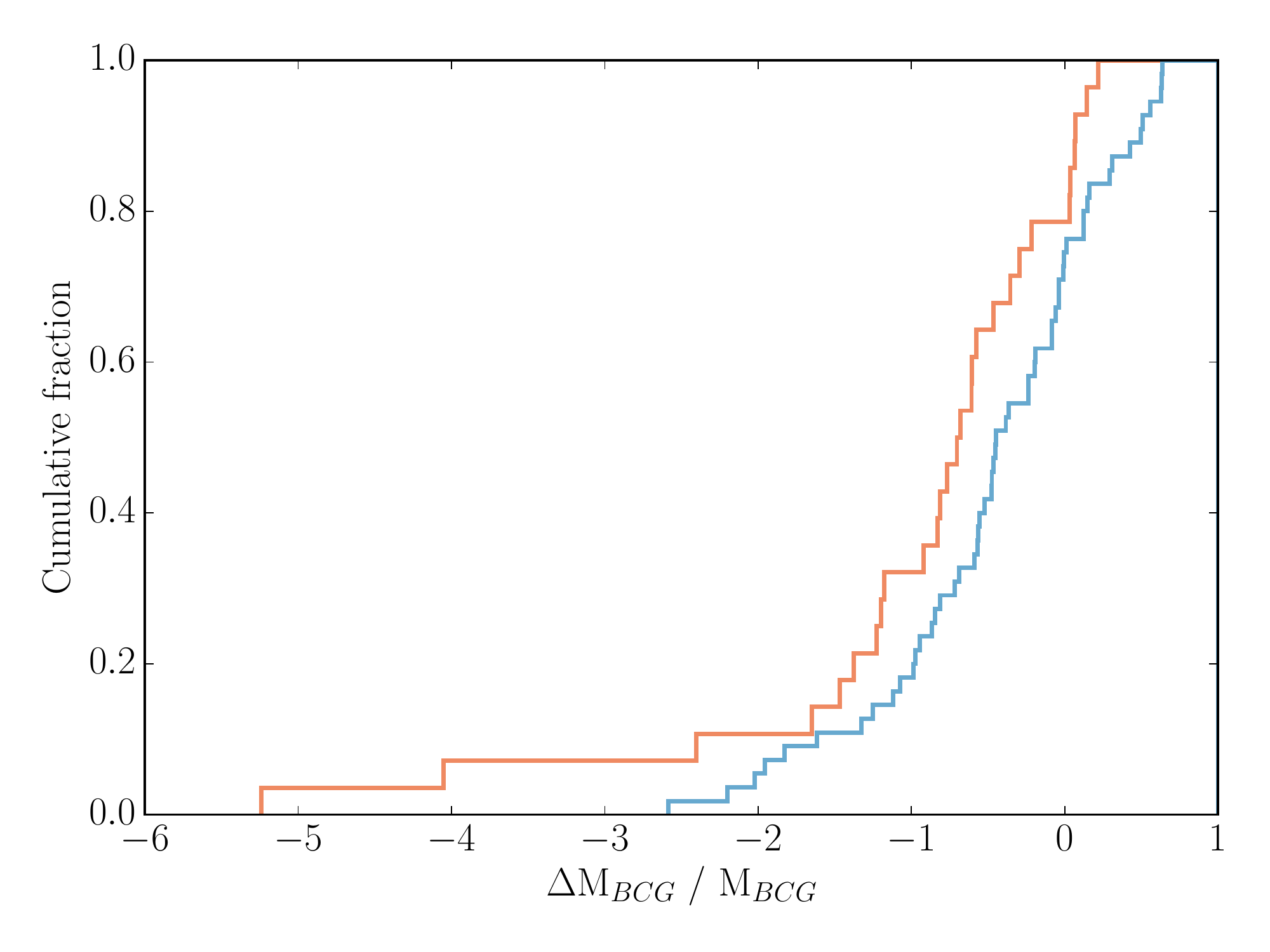}
\caption{Cumulative normalized BCG~$\Delta$M for offsets greater than
  0.05$\times r_{500}$ (red) and smaller than 0.05$\times r_{500}$
  (blue). }
\label{bcg_delmass_cumul}
\end{figure}

Figure~\ref{bcg_delmass_cumul} shows the cumulative distribution of
BCGs mass lag measurement $\Delta M/M_{BCG}$. A Shapiro test for
normality reveals that BCGs located in relaxed clusters are normally
distributed (although the mean is not zero) while the BCGs located in
unrelaxed clusters are not and follow more closely a log-normal
distribution. To compare the distributions, we employ an
  Anderson-Darling test. This nonparametric test is used to assess
  wether or not two samples come from the same distribution by
  computing the maximum deviation between their cumulative
  distribution. It is very similar to the Kolmogorov-Smirnov test but
  differs in that it is better suited to samples with different mean
  values or outliers. We find that we can reject the null hypothesis
that the samples are drawn from the same distribution with 95\% confidence (\emph{p-value}$<0.05$), a value that goes up to 99.6\% (\emph{p-value}$<0.004$) if 
we compare clusters with offset lower than 0.05$\times r_{500}$ to ones 
with offsets greater than 0.1$\times r_{500}$.

\subsection {$\Delta$m$_{12}$ and the BCG merger history}
\label{sec_del_m12}

The luminosity gap between the first and second brightest cluster
members, $\Delta$m$_{12}$, provides a measure of cluster galaxy
evolution.  The hierarchical accretion model of galaxy growth predicts
that the BCG within a cluster should grow in mass faster than
non-central, non-dominant galaxies as the BCG is located at the centre
of the cluster potential to which less massive galaxies migrate via
dynamical friction.  Therefore, if BCGs grow via such accretion, one
expects the luminosity gap to grow with every accreted galaxy
\citep[e.g.][]{locuss_m12,four_d_class}. Cluster-scale merger events 
can affect the evolution of $\Delta$m$_{12}$ as they can add bright galaxies, 
reducing the luminosity gap.

To compute $\Delta$m$_{12}$ for each cluster we first define cluster
membership.  Since have we only photometric redshifts for most non-BCG
galaxies, we put stringent constraints on the membership
classification to reduce contamination. We define a galaxy as a member
of a given cluster if the cluster redshift is within the 1-$\sigma$
range of the photometric redshift of the galaxy and the galaxy lies
within $1\times r_{500}$ of the X-ray centroid. If the galaxy has a
spectroscopic redshift, it is used instead of the photometric
redshift. A galaxy then has to be within 3000~km~s$^{-1}$ of the
cluster to be considered a member. We then set the value of
$\Delta$m$_{12}$ as the difference in \emph{z}-mag between the BCG and
the second brightest member.

We apply a Spearman rank correlation test to determine the extent to
which $\Delta$m$_{12}$ is correlated with measurements of BCG mass and
mass-lag $\Delta M/M_{BCG}$ across the XXL-100-GC sample.  Noting that
one also expects $\Delta$m$_{12}$ to increase with time, we compute
correlation values correcting for any partial correlation with
redshift according to the formula:
\begin{equation*}
S_{AB|C}=\frac{S_{AB} - S_{AC}S_{BC}}{\sqrt[]{(1 - S_{AC}^{2})(1 - S_{BC}^{2})}},
\end{equation*}
where $S_{AB|C}$ is the Spearman rank correlation between A and B,
corrected for C. This test indicates that $\Delta$m$_{12}$ is
correlated positively with BCG mass and $\Delta M/M_{BCG}$ at 99.5\% and 99.8\%
confidence level respectively and that $\Delta$m$_{12}$ is a reliable
tracer of BCG mass growth. Individual values of $\Delta$m$_{12}$ can
be found in Table~\ref{bcg_position}.

The value of $\Delta$m$_{12}$ does not indicate the mass distribution
of accreted galaxies.  To address this question we consider the
results on BCG growth taken from the Millennium Simulation
(\citealt{springel_mil}) at $z<1$ using the \texttt{DeLucia2006a}
semi-analytical galaxy models data presented in
\cite{delucia_blaizot_2007}. These models were obtained from the
Millennium database\footnote{http://gavo.mpa-garching.mpg.de/portal/}
(\citealt{lemson_mil}).  One hundred BCGs were randomly selected at
$z=0$ from clusters with M$_{cluster}\sim 2\times10^{14}$M$_{\odot}$,
i.e. the average XXL-100-GC cluster mass. For each BCG we obtain the
merger tree between $z=1$ and $z=0$ (\citealt{lemson_tree}).

\begin{figure}
\centering
\includegraphics[width=0.99\columnwidth]{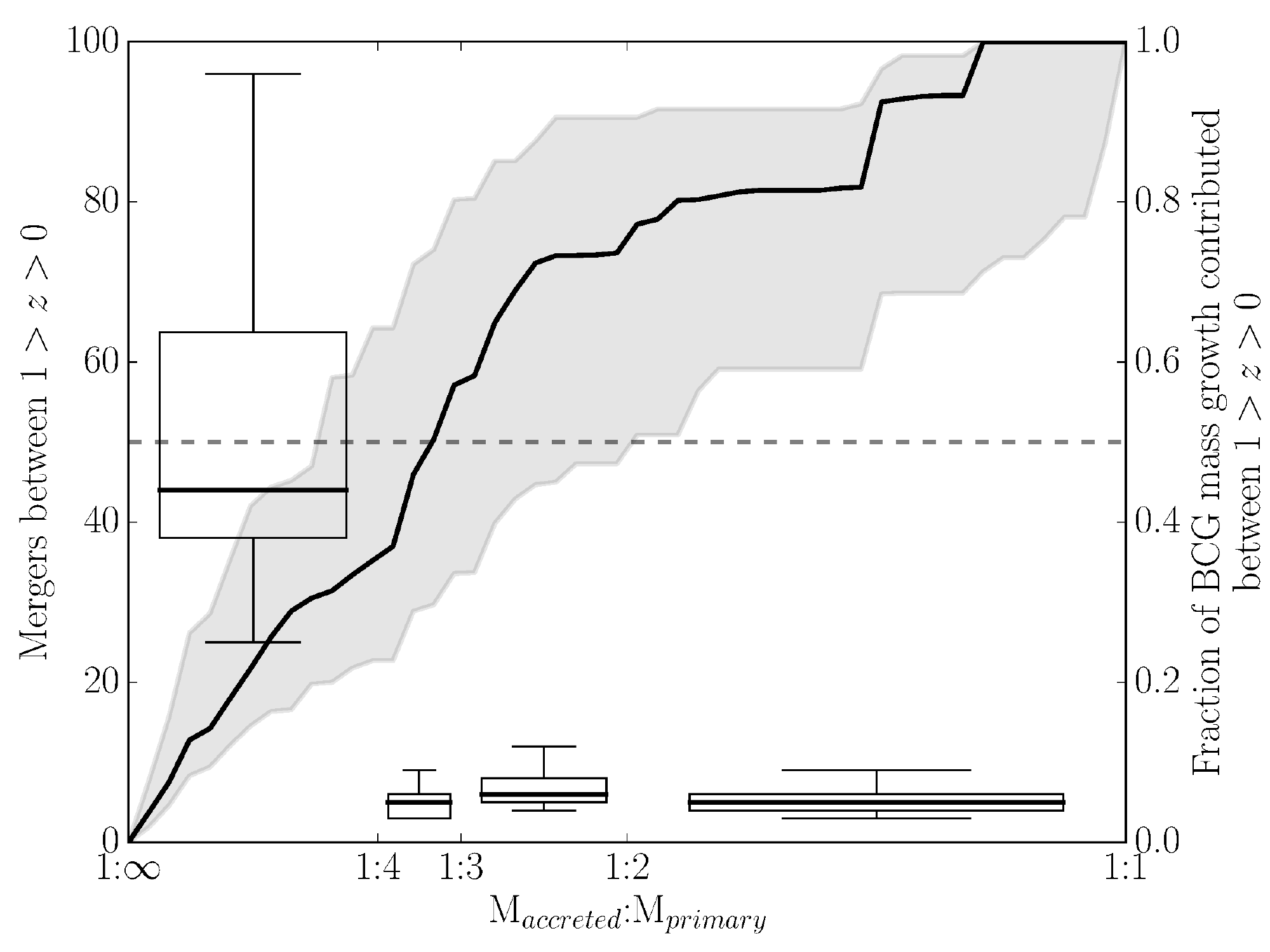}
\caption{Accretion history over $0<z<1$ obtained for 100
  semi-analytical BCG galaxy models realised within
  \texttt{DeLucia2006a}.  The rectangular boxes indicate the number of
  mergers for mass ratio intervals of $1:\infty -1:4$, $1:4 - 1:3$,
  $1:3 - 1:2$ and $1:2 - 1:1$.  The top and bottom of each box marks
  the upper and lower quartile values while the interior horizontal
  line indicates the median value.  The error bars indicate the 5-th
  and 95-th percentiles.  The solid black line indicates the median
  value of the normalized cumulative merger contribution to BCG mass
  growth.  The accompanying shaded grey region indicates the full
  extent of the 100 normalized cumulative mergers to the BCG mass
  growth.  The grey horizontal dashed line indicates the point at
  which the BCG has accreted 50\%\ of its $z=0$ mass.}
\label{millenium_bcg}
\end{figure}

Figure~\ref{millenium_bcg} shows the distribution of the number of
mergers in bins of mass ratio for the 100 BCGs in addition to the
fractional contribution to the $z=0$ BCG mass.  From the Figure it is
clear that one-half of the $z=0$ BCG mass is each contributed from
mergers at mass ratios greater than and less than a value of $1:3$.
Although there is a certain amount of scatter about the mean
relationship displayed in Figure \ref{millenium_bcg}, the results from
simulations appear to be in broad agreement with those of
\cite{burke_bcg} obtained with \emph{HST} imaging of BCGs and their
bound companions around $z\sim1$.

\subsection[]{Luminosity segregation}
\label{sec_segregation}

An alternative diagnostic of the hierarchical accretion of cluster
galaxies is to consider their luminosity segregation. A prediction of
this hypothesis is that the central regions of a galaxy cluster should
be overabundant in bright galaxies relative to faint as brighter
(i.e. more massive) galaxies are expected to migrate faster to the
cluster centre under the influence of dynamical friction. One further
expects that this overabundance of bright galaxies will be more marked
in relaxed clusters compared to those which are unrelaxed.

The luminosity segregation method proposed by
\cite{lidman_segregation} compares the cumulative spatial distribution
of bright galaxies to faint ones. They employ a two sample KS
  test on the two distributions and find a significant difference in
  the radial distribution of faint and bright galaxies yet note that
  this result is very sensitive to the arbitrary maximum radius to
  which the calculation is performed. Unlike
  \citeauthor{lidman_segregation}, we know the value of $r_{500}$ for
  all the clusters and use it as the maximum radius.  Although still
arbitrary, the use of $r_{500}$ as the maximum radius used in the same
calculation applied to the XXL-100-GC sample does at least provide a
consistent and physically-motivated maximum radius for each cluster.

\begin{figure}
\centering
\includegraphics[width=0.99\columnwidth]{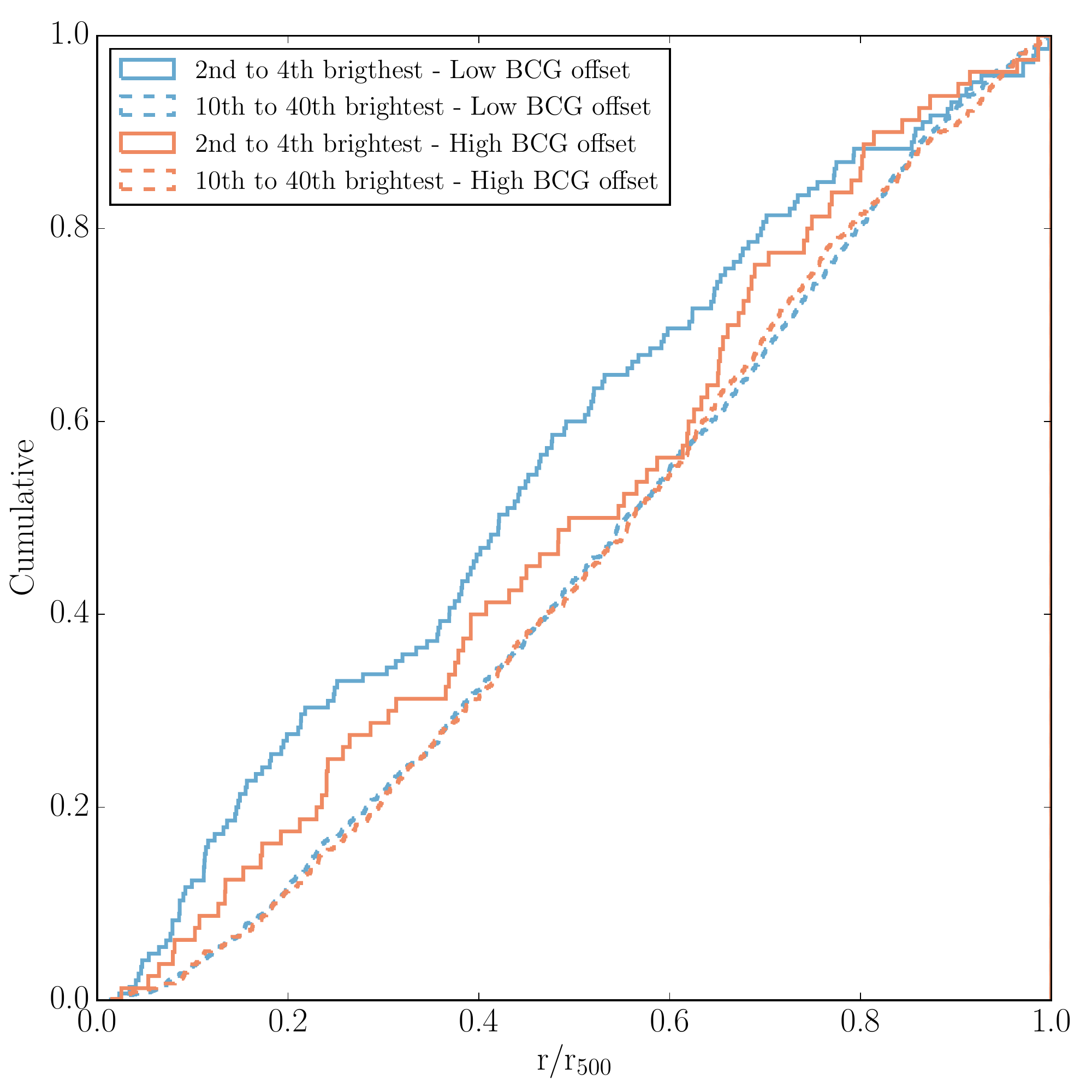}
\caption{The cumulative radial distribution of bright and faint
  galaxies in relaxed ($ <0.05 \times r_{500}$) and unrelaxed ($>0.05
  \times r_{500}$) clusters.  The solid blue and red lines shows the
  distribution of 2nd to 4th brightest members in relaxed and
  unrelaxed clusters.  The dashed blue and red lines show the
  distribution of 10th to 40th brightest galaxies in relaxed and
  unrelaxed clusters.}
\label{mass_seg}
\end{figure}

We compare the cumulative radial distribution of bright and faint
cluster members in XXL-100-GC clusters. We define bright galaxies as the
2nd, 3rd and 4th brightest members. Faint galaxies are defined as the
10th to 40th brightest members. Figure~\ref{mass_seg} shows the
cumulative distributions of faint and bright galaxies within $r_{500}$
for the relaxed and unrelaxed clusters. A two-sample
  Anderson-Darling test reveals that the radial distribution of bright
  galaxies in unrelaxed clusters (\emph{red}) is not very different
  than that of faint ones (\emph{p-value}$=0.13$, 30 clusters). However, in relaxed clusters (\emph{blue}) a
  significant difference exists (\emph{p-value}$=1.97\times10^{-5}$,
  55 clusters). 
 
\subsection[]{H$\alpha$ star formation}
\label{sec_ha}

Brightest cluster galaxies typically appear as passively evolving
stellar populations. However, observed stellar masses grow by a factor
$\sim 2$ between $z=1$ and the present epoch \citep{lidman_mass}.  
Active star formation in BCGs is observed and in the
  literature has been interpreted as evidence for inflows of cool gas
  within the cluster potential (e.g. \citealt{donahue_1992}; \citealt{edge_1992}; \citealt{odea_2010}).
Evidence of active star formation associated with an infall of gas
from cooling flows is also observed by \cite{locuss_sf} in the spectra of
some BCGs in the LoCuSS sample.  We assess the presence of active star
formation in the sample of XXL-100-GC BCGs by focussing on a sub-sample
of 30 BCGs in the Northern field with $z\lesssim0.5$ for which
H$\alpha$ emission fluxes have been measured by the SDSS Portsmouth
group from dust extinction-corrected DR12 data (\citealt{sdss_em}).

Although the spectra have high signal-to-noise ratio (SNR), only half
of the BCGs show H$\alpha$ emission detection with a line
SNR~$\gtrsim2$ down to an observed flux of $\sim 1 \times 10^{-17}$ erg~cm$^{-2}$~s$^{-1}$. 
Using the classical
  H$\alpha$ flux to SFR conversion from \cite{osterbrock} and the
  stellar masses we determined in Section~\ref{sec_bcg_mass}, we
  confirm that none of the $z\lesssim0.5$ BCGs in XXL-N shows a sSFR
  greater than $\sim10^{-12}$~yr$^{-1}$.  While the exact value of the
  star formation rate expected for a passive BCG is unclear,
  observations and simulations provide some
  guidance. \cite{zwart_star_form} use 1.4GHz VLA data from a $K_S$
  selected sample of galaxies in the VIDEO survey to deduce a sSFR of
  $\sim10^{-11}$~yr$^{-1}$ for $\sim10^{11}~[M_{\star}/M_{\odot}]$
  elliptical galaxies with $0<z<1$.  \cite{henriques_star_form} find a
  sSFR of $\sim10^{-12}$~yr$^{-1}$ for similar masses and redshift in
  simulations.

We therefore conclude that none (less than 3\%) of
  $z<0.5$ XXL-100-GC BCGs display evidence for enhanced star formation
  above that expected for field ellipticals of comparable mass and
  redshift. It is important to note that we do not possess any
  spectroscopic emission line constraint on the current SFR in $z>0.5$
  XXL-100-GC BCGs.

\subsection[]{Red sequence offset}
\label{sec_red_seq} 

\begin{figure}
\centering
\includegraphics[width=0.99\columnwidth]{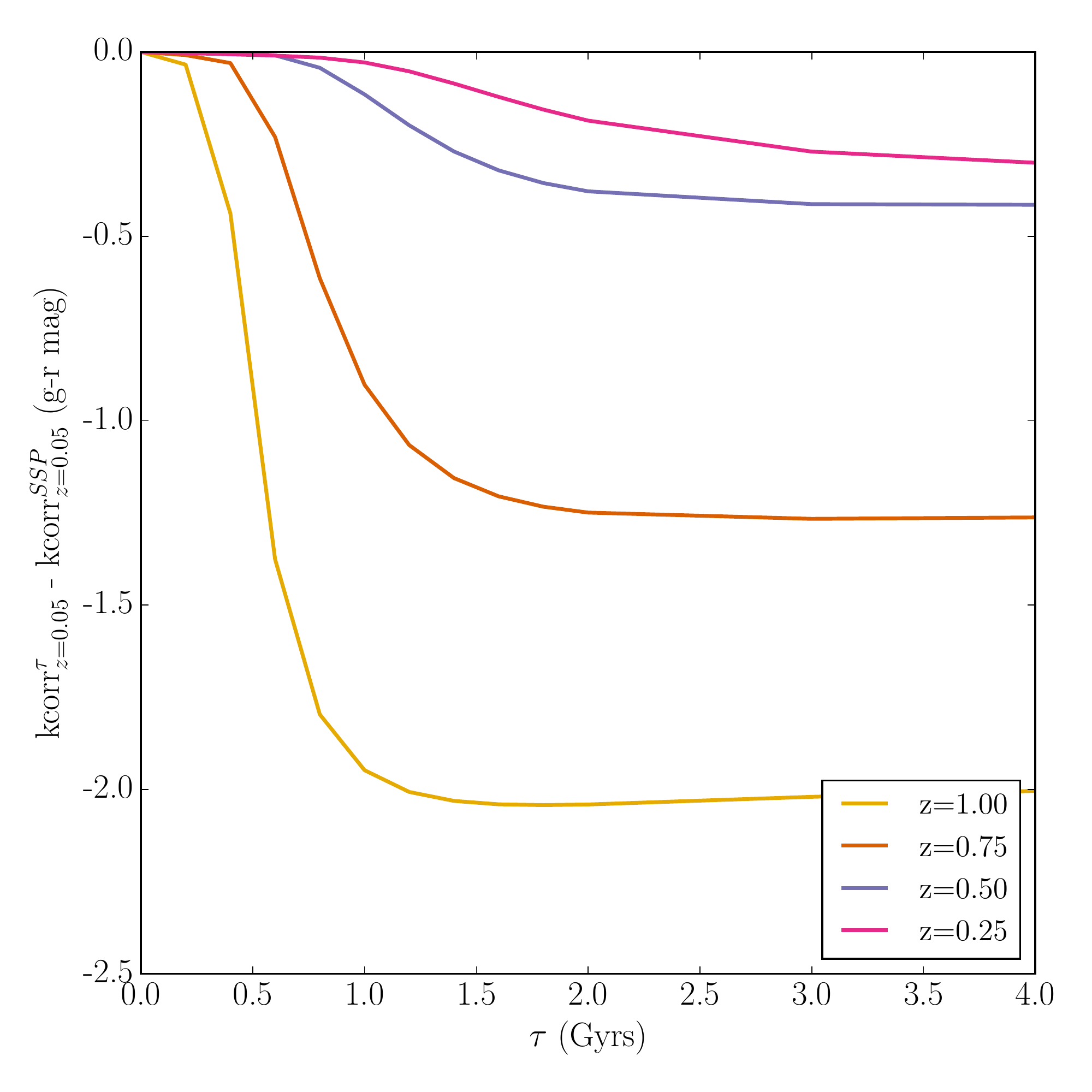}
\caption{The difference in $k$-correction required to correct a galaxy
  at a specified observed redshift to $z=0.05$ assuming a SSP and an
  exponentially decreasing star formation rate of timescale $\tau$.}
\label{kcorr_tau}
\end{figure}

The analysis of XXL-100-GC BCG stellar masses (Section
  \ref{sec_bcg_mass}) and H$\alpha$ emission line fluxes at
  $z\lesssim0.5$ (Section \ref{sec_ha}) indicate that these BCGs
  display low specific star formation rates. However, we note that
  because of a combination of the wavelength coverage of SDSS
  spectroscopy and the fact that the majority of BCGs are located at
  $z<0.5$, these tests are weighted towards the properties of low
  redshift BCGs within the XXL-100-GC sample. There are 19 BCGs at $z >
  0.5$, meaning we are left uninformed on possible star formation in a
  fifth of the sample. This section tries to address this with an
  alternative test of star formation in XXL-100-GC BCGs based on the
  magnitude of BCG colour offsets from their host cluster red
  sequences. As Figure~\ref{kcorr_tau} indicates, the $k$-correction
applied to galaxies at greater redshift is more sensitive to
deviations from the assumption that BCG spectra are described by an
old, passively evolving single stellar population (SSP).

\begin{figure}
\centering
\includegraphics[width=0.99\columnwidth]{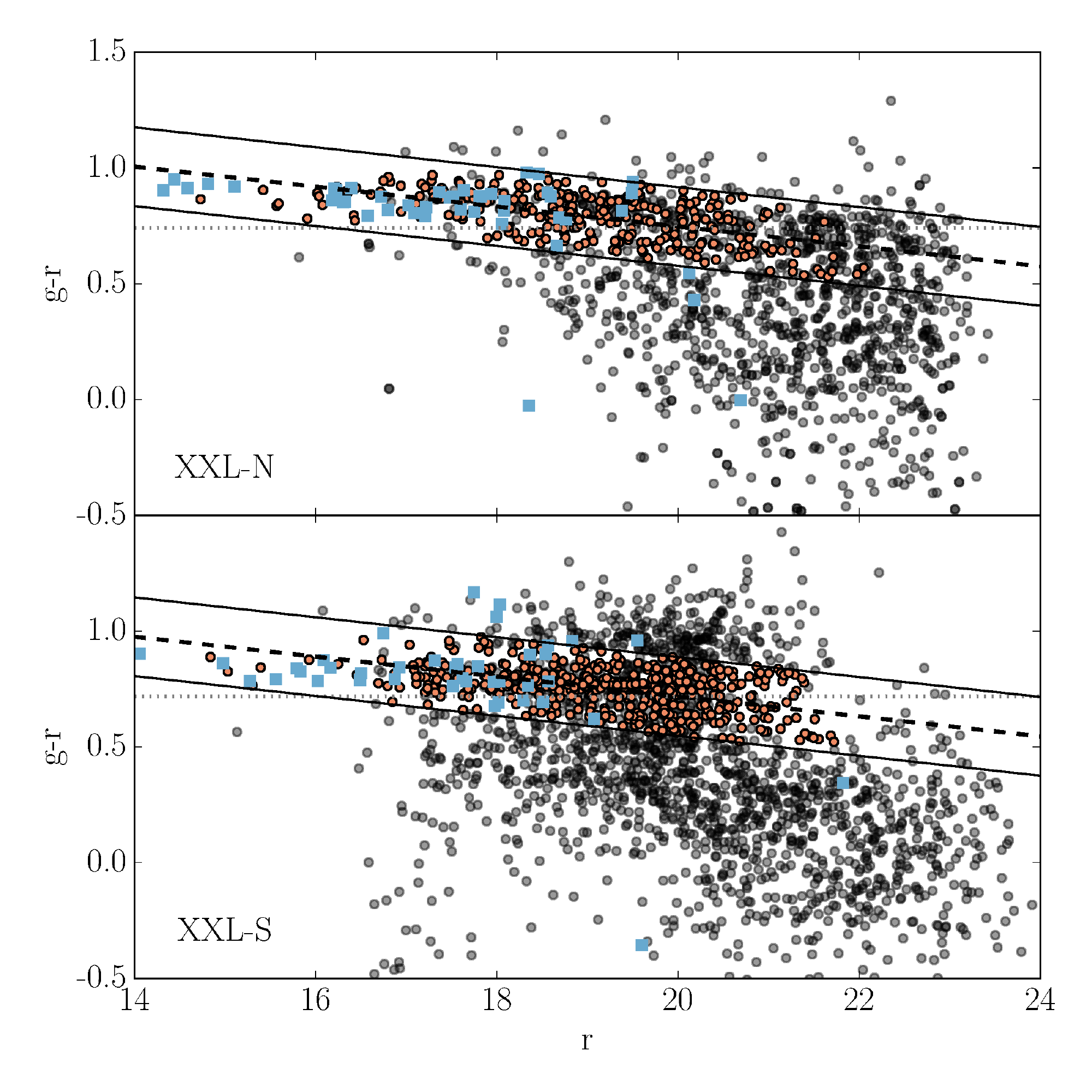}
\caption{The colour-magnitude diagram of all XXL-100-GC North (\emph{top
    panel}) and South (\emph{bottom panel}) member galaxies
  $k$-corrected to $z=0.05$ considering a SSP star formation history.
  The grey dotted lines show the initial red sequence lower colour
  limit obtained from the double Gaussian fit.  The solid black lines
  show the converged red sequence colour cutoff.  The red dots
  indicate all red sequence galaxies with M$_{V}\leq-20$; The blue
  squares indicate BCGs.  The black dashed lines show the best fitting
  red sequence relation.}
\label{red_seq}
\end{figure}

We separately create a single stacked colour magnitude diagram for all
cluster member galaxies located within each of the XXL-N and XXL-S
fields.  We apply a $k$-correction based upon the best-fitting star
formation history obtained in Section \ref{sec_bcg_mass} and a
distance modulus correction to stack all member galaxy photometry at
an assumed $z=0.05$.  Member galaxies are selected employing the
criteria outlined in Section~\ref{sec_del_m12}.  We determine the
location of the stacked cluster red sequence on each colour magnitude
diagram employing an iterative process.  Firstly, considering only
galaxies with M$_{V}\leq-20$, we fit a simple double Gaussian
distribution to the colour distribution of member galaxies and define
the red sequence cutoff as the color at which the contribution of blue
and red galaxies are equal.  We then fit a linear red sequence from
those galaxies redder than this cutoff and, using the
$\Delta(B-V)=-0.2$ criterion for blue galaxies from \cite{bo_effect},
we refine the red sequence by re-selecting red galaxies as the ones
for which $g-r$ colour falls within $\Delta(B-V)=\pm0.2$ of the linear
fit.  This process is repeated until it converges and the slope in
XXL-S is fixed to be the same as the one in XXL-N. Doing so makes the
red sequence in XXL-S slightly steeper but limits the contribution of
the large number of dubious $g-r~>1.0$ galaxies in the field that may
be caused by the larger photometric errors in this field.
Figure~\ref{red_seq} shows the resulting colour-magnitude diagram
(corrected to SDSS $g-r$) of all member galaxies from both fields
after $k$- and distance modulus correction.

The resulting distribution of BCG offsets from the stacked red
sequence in each field is consistent with a Gaussian distribution of
zero mean.  In both XXL-N and XXL-S, the distribution has a standard
deviation $\sigma (g-r) \approx 0.07$: a relatively small deviation
that indicates that most BCGs lie close to the red sequence.  It is
perhaps no surprise that the XXL-100-GC BCGs lie at low colour offset
from the red sequence: these represent the bulk of the systems for
which we have good quality spectroscopy and to which the SFH analysis
applied in Section \ref{sec_bcg_mass} is most sensitive.  

As mentioned previously, earlier analyses indicate that XXL-100-GC BCGs
have passively evolving SFHs.  However, Figure~\ref{kcorr_tau}
indicates that the $k$-correction applied to transform a BCG at
$z>0.5$ to the $z=0.05$ colour-magnitude plane is very sensitive to
deviations from an assumed old, co-eval SSP model.  SSP models
computed assuming $\tau \le 1$~Gyrs fall within our $3-\sigma$
confidence limits displayed in Figure~\ref{SFH-contours}.  One can
therefore employ the absence of BCGs with $z>0.5$ and large colour
offsets from the stacked red sequence as evidence that these systems
are also consistent with SSP models possessing $\tau \le 1$~Gyrs.  In
fact, out of the 19 BCGs at $z>0.5$, we find only one with an offset
that can only be explained with a $\tau\gtrsim1$~Gyrs: XLSSC~546.  It
is unfortunate that this system lacks a spectroscopic redshift which
might indicate the presence of active star formation.  However, a
closer inspection of the X-ray contours of XLSSC~546 reveals that the
BCG sits within one of two X-ray peaks observed in the cluster,
suggesting the cluster is disturbed and possibly experiencing a merger
event.

\section{Discussion}
\label{sec_discussion}

\begin{figure*}
\centering
\includegraphics[width=0.99\textwidth]{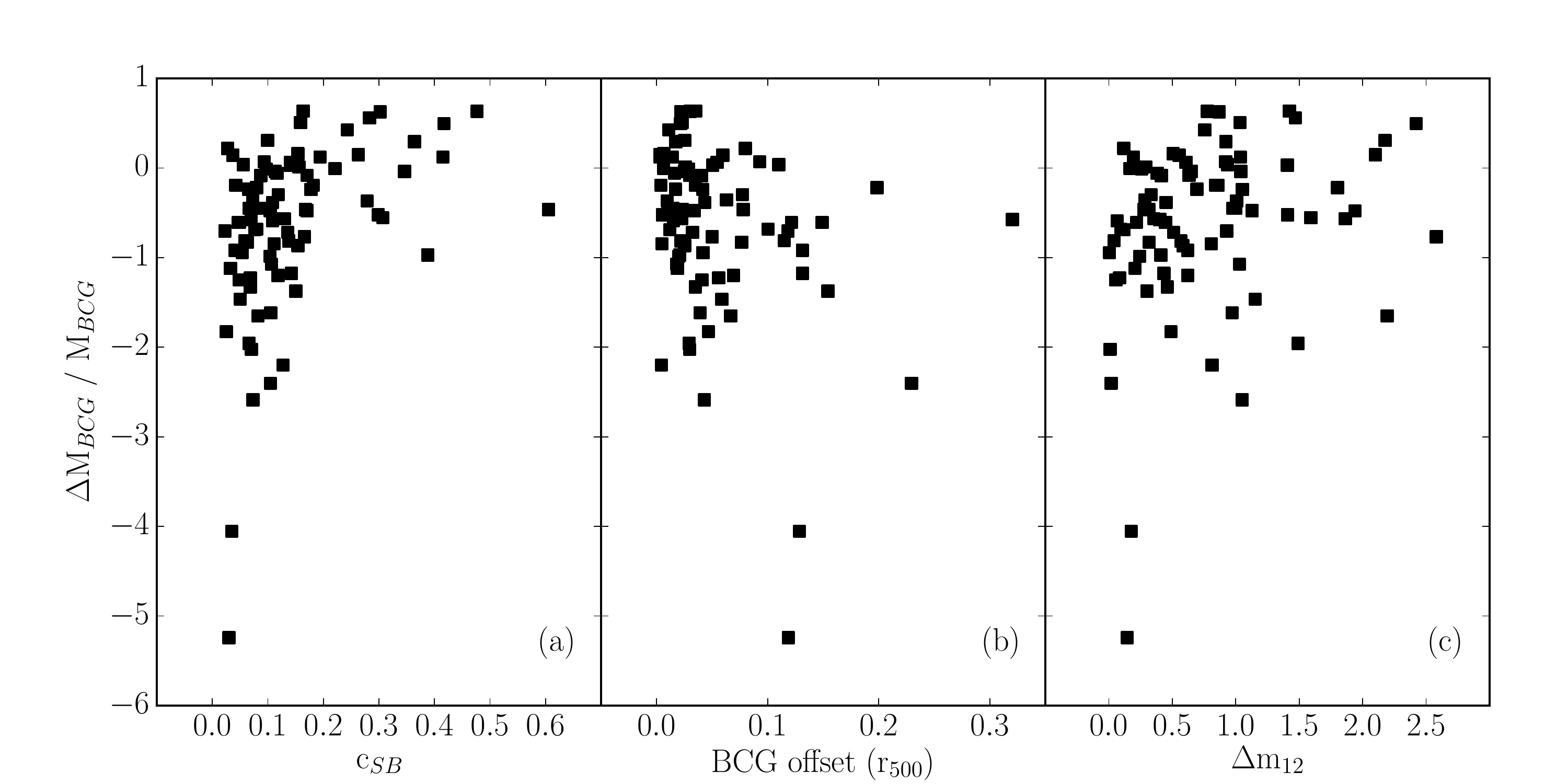}
\caption{Normalized $\Delta$M$_{BCG}$ for various indicators. The
  value of $\Delta$M$_{BCG}$ is the difference between a BCG observed
  mass and the \emph{expected} mass for a BCG in a relaxed cluster of
  the same mass obtained from the M$_{cluster}$ - M$_{BCG}$
  relation. (a) Cluster relaxation from c$_{SB}$. (b) Dynamical
  relaxation from BCG offset. (c) $\Delta$m$_{12}$ tracing BCG
  accretion.}
\label{bcg_delmass_relations}
\end{figure*}

We have determined that, within the sub-sample of relaxed XXL-100-GC
clusters, the BCG stellar mass is linearly related to the cluster weak
lensing mass. We compute a value of $n=1.04\pm0.20$ and $1.03\pm0.10$
respectively for the power-law index of the M$_{cluster}$ - M$_{BCG}$
relation for XXL-100-GC clusters which appear relaxed either via their
$c_{SB}$ weighting or based upon low BCG offset ($<0.05\times
r_{500}$). These index values are generally lower
  than reported in the literature and may be due to three
  considerations: 1) the XXL-100-GC sample extends to lower mass
  clusters, 2) we explicitly differentiate between relaxed and
  unrelaxed systems and 3) flux-selected samples like XXL-100-GC and
  HIFLUGCS contain a larger fraction of disturbed systems compared to
  luminosity selected cluster samples. Lower cluster mass correlates
with lower member galaxy velocity dispersions \citep{willis_2005}.  As
the cluster velocity dispersion approaches that of the BCG, the
effective merger cross section increases rapidly
\citep[e.g.][]{makino1997}.  This assertion is supported by various
analysis (e.g.: \citealt{gonzales_2007}; \citealt{leauthaud_2012};
\citealt{coupon_2015}; \citealt{xxl_x}) that indicate that BCGs
contribute a greater fraction of the total cluster stellar luminosity
in lower mass clusters, as expected if stellar mass is more
efficiently accreted by the BCG.

Perhaps more important than the exact value of the slope of the
M$_{cluster}$ - M$_{BCG}$ relation is the result that relation is
statistically different for relaxed and unrelaxed clusters. The
relation for clusters with a disturbed BCG is much shallower at
$n=0.55\pm0.16$ and $0.46\pm0.17$ respectively for clusters weighted
by inverse $c_{SB}$ or for large BCG offset ($>0.05\times r_{500}$).
This indicates that, when a cluster gains mass via a merger, the BCG
stellar mass initially lags behind the value expected for a dominant
galaxy in a cluster with the mass of the merged host. The effect of 
a cluster-scale merger is therefore more readily detectable
via the increased cluster mass (inferred from the ICM temperature)
rather than the stellar mass of the BCG. 

Although star formation in BCGs can be caused by the infall of gas
from cooling flows, XXL-100-GC clusters display low
  central gas concentrations.  
Within XXL-100-GC we have used spectroscopic observations of the
H$\alpha$ line as a direct star formation indicator for a third of the
sample. We find no H$\alpha$-determined sSFRs above the value observed
in similar mass, passive galaxies in the field.

Furthermore, the analysis of BCG offsets from the global cluster red
sequence indicates that only one high redshift BCG in XXL-100-GC, located
in a potentially merging cluster, shows evidence for a stellar
population described by a declining star formation rate of timescale
$\tau\gtrsim1$~Gyrs.  In fact, the almost complete absence of active
star formation observed in the BCG population motivates our choice of
a single stellar population model to describe the SFH of XXL-100-GC BCGs.
The population of XXL-100-GC BCGs therefore appears to be homogeneously
passive irrespective of the relaxation state of the parent cluster.
This realisation is in agreement with results from \cite{webb_2015}
and \cite{mcdonald_2015} indicating that dry mergers are the dominant
source of growth in BCGs at $z\lesssim1$. Another important
  factor at play is that XXL-100-GC clusters are less massive on average
  than their LoCuSS and CLASH counterparts. \cite{liu_2012} show that
  the incidence of star formation in BCGs increases with cluster
  richness and X-ray luminosities, both cluster mass proxies. In
  agreement with what we report in this work, XXL-100-GC clusters should
  host BCGs with significantly lower star formation on average than
  those in the LoCuSS and CLASH sample.

Figure~\ref{mass_seg} indicates that bright galaxies
  have a dominant contribution at low radii in clusters with a BCG
  offset of $<0.05\times r_{500}$. In this case, an Anderson Darling
  test between bright and faint galaxies indicates that we can exclude
  that they come from the same distribution at $>$99.99\%. The same
  test applied clusters with a BCG offset of $>0.05\times r_{500}$
  cannot exclude the null hypothesis. The test suggests that, as the
  cluster evolves, so does the galaxy distribution. This is important
  as such infalling bright galaxies could present a major source of
  BCG stellar mass growth via major mergers as they contribute
  typically half of the BCG growth according to~\cite{burke_bcg} and
  our results from Section~\ref{sec_del_m12}. However, we note that
  the statistical significance varies according to what we define as a
  \emph{bright} or \emph{faint} galaxy. Nevertheless, the results
  generally indicate the presence of mass segregation.

In Figure~\ref{bcg_delmass_relations}, we attempt to combine a number
of observational measures to generate an overview of BCG evolution in
galaxy clusters.  The leftmost panel of
Figure~\ref{bcg_delmass_relations} reveals that we observe no XXL-100-GC
BCGs with a high mass lag (negative values) in clusters where the
X-ray gas is very relaxed.  The BCG is clearly gaining stellar mass
and reducing the inferred mass lag before the bulk of the X-ray gas
can settle in the cluster potential.  This point is relevant as the
XXL-100-GC BCGs show essentially no evidence for active star formation.
This in turn indicates an absence of significant gas accretion as the
gas remains disturbed on timescales longer than stellar mass accretion
to the BCG.  

The middle panel of Figure~\ref{bcg_delmass_relations} shows that the
BCG grows in stellar mass relative to the total cluster mass as the
BCG moves toward the centre of the X-ray emitting gas (which we interpret
as the centre of the cluster potential). A range of trajectories
appear to converge toward the upper left corner of the diagram (zero
mass lag), indicating a certain amount of scatter in the stellar mass
growth history of individual BCGs.  However, despite this scatter, the
absence of points in the lower left region of the diagram indicates
that there exist no relaxed clusters in which the BCG displays a
significant mass lag.

The right panel in Figure~\ref{bcg_delmass_relations} indicates that
the stellar mass in the BCG grows relative to the second brightest
cluster galaxy (a similar trend is observed whether one employs the
2nd, 3rd or 4th brightest galaxy as a reference) as it also grows
relative to the total cluster mass.  The analysis of luminosity
segregation contained in Section \ref{sec_segregation} indicates that
bright galaxies in relaxed clusters are preferentially located at low
cluster centric radius compared to both bright galaxies in unrelaxed
clusters and faint galaxies in all clusters.  We interpret this result
as the effect of dynamical friction operating undisturbed in relaxed
clusters.  The accretion of such bright, infalling galaxies onto
XXL-100-GC BCGs provides a compelling statistical explanation for the
trend of $\Delta m_{12}$ versus mass lag shown in the right panel of
Figure~\ref{bcg_delmass_relations} and appears to agree well with
simulations which indicate that major mergers might contribute 50\% on
average of the stellar mass growth in BCGs at $z<1$.

Finally, BCGs at low-, intermediate- and high-\emph{z} all broadly
cover the same regions of Figure~\ref{bcg_delmass_relations}.  This
impression can be further verified by the application of a Spearman
rank correlation test. For all indicators ($c_{SB}$, BCG offset and
$\Delta m_{12}$), we find no significant difference in the correlation
with BCG mass lag when performing a regular test compared to a partial
test correcting for redshift.  This would appear to indicate that,
although the merger rate of clusters may vary in a secular fashion
with cluster mass and redshift, the physical response of the BCG to
these stochastic events is independent of redshift.

\section*{Conclusion}

\begin{figure}
\centering
\includegraphics[width=0.99\columnwidth]{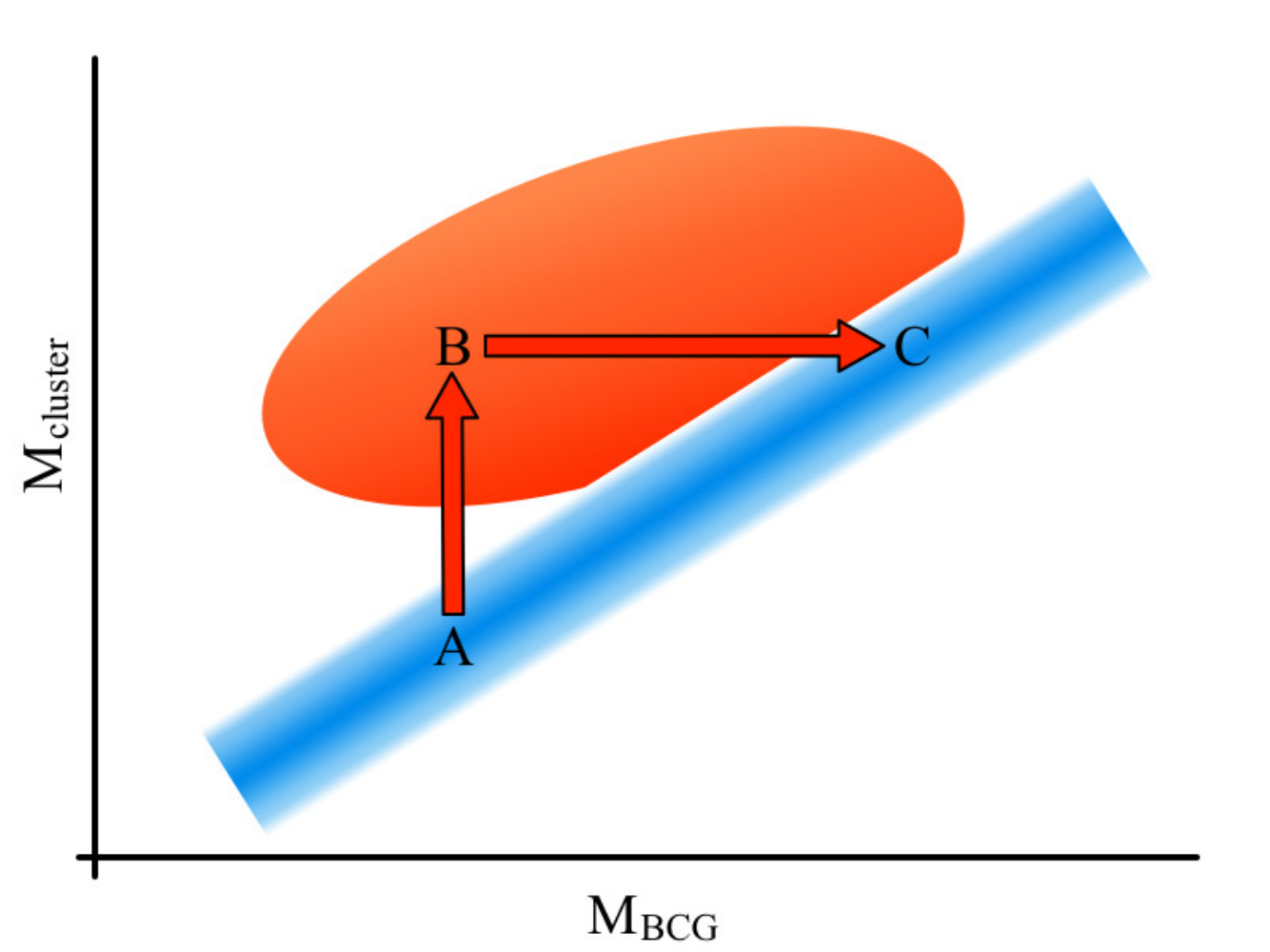}
\caption{Cartoon of the BCG mass growth through dry merger in XXL-100-GC. The \emph{blue} region represent the expected relation for relaxed clusters while the \emph{red} region is where disturbed clusters are found due to their BCG mass lag.} 
\label{bcg_growth}
\end{figure}

The story told by XXL-100-GC can be summarized by a cartoon presented in
Figure~\ref{bcg_growth}.  In this scenario, an idealized cluster is
initially relaxed and the BCG mass is such that it lies at point A, in
agreement with the relationship $\rm M_{BCG} \propto M_{cluster}$.
Following a cluster-scale merger event, the cluster mass increases and
the ICM of the merged cluster is shock heated to the virial
temperature of the new system.  Any cool core system present is
disrupted and the BCG is displaced from the centre of the
cluster potential.  At this moment, the system is located at point B
in Figure \ref{bcg_growth}: the ICM temperature reflects the total
mass of the system but the BCG now lags in mass relative to the
cluster.  As the cluster begins to relax, the BCG and other bright
galaxies preferentially migrate to the cluster centre under the
influence of dynamical friction.  These bright galaxies ultimately
merge with the BCG, both increasing the BCG stellar mass relative to
the cluster and increasing the value of $\Delta m_{12}$.  At this
instance in time the cluster approaches point C on Figure
\ref{bcg_growth}.

Despite the outline above several questions remain: Can the rate of
BCG stellar mass accretion be quantified by searching for
morphological evidence of merging in high-spatial resolution images of
BCGs \citep[e.g.][]{liu_2015}?  In addition, how does the relationship
between M$_{cluster}$ and M$_{BCG}$, which is observed to steepen in
cluster samples of greater mass (\citealt{stott_2010};
\citealt{lidman_mass}; \citealt{stott_2012}), depend upon the inferred
relaxation state? At what cluster mass does cooling-flow induced BCG
star formation become an important mechanism for BCG stellar mass
growth \citep[e.g.][]{sanderson2009}?  A sensible extension to this
work would therefore be to study the properties of BCG mass lags in a
sample of clusters of higher typical mass than XXL-100-GC.

\section*{Acknowledgments}
XXL is an international project based around an \emph{XMM-Newton} Very Large Programme surveying two 25 deg$^{2}$ extragalactic fields at a depth of $\sim5 \times 10^{-15}$~erg~cm$^{-2}$~s$^{-1}$ in the [0.5-2] keV band for point-like sources. The XXL website is http://irfu.cea.fr/xxl. \\
This paper uses data from observations obtained with MegaPrime/MegaCam, a joint project of CFHT and CEA/IRFU, at the Canada-France-Hawaii Telescope (CFHT) which is operated by the National Research Council (NRC) of Canada, the Institut National des Science de l'Univers of the Centre National de la Recherche Scientifique (CNRS) of France, and the University of Hawaii. This work is also based in part on data products produced at Terapix available at the Canadian Astronomy Data Centre as part of the Canada-France-Hawaii Telescope Legacy Survey, a collaborative project of NRC and CNRS. \\
This research uses data from the VIMOS VLT Deep Survey, obtained from the VVDS database operated by Cesam, Laboratoire d'Astrophysique de Marseille, France. \\
Based in part on data acquired through the Australian Astronomical Observatory \\
This paper uses data from the VIMOS Public Extragalactic Redshift Survey (VIPERS). VIPERS has been performed using the ESO Very Large Telescope, under the "Large Programme" 182.A-0886. The participating institutions and funding agencies are listed at http://vipers.inaf.it. \\
GAMA is a joint European-Australasian project based around a spectroscopic campaign using the Anglo-Australian Telescope. The GAMA input catalogue is based on data taken from the Sloan Digital Sky Survey and the UKIRT Infrared Deep Sky Survey. Complementary imaging of the GAMA regions is being obtained by a number of independent survey programmes including GALEX MIS, VST KiDS, VISTA VIKING, WISE, Herschel-ATLAS, GMRT and ASKAP providing UV to radio coverage. GAMA is funded by the STFC (UK), the ARC (Australia), the AAO, and the participating institutions. The GAMA website is http://www.gama-survey.org/. \\
Funding for SDSS-III has been provided by the Alfred P. Sloan Foundation, the Participating Institutions, the National Science Foundation, and the U.S. Department of Energy Office of Science. The SDSS-III web site is http://www.sdss3.org/. \\
Based in part on data acquired through the Australian Astronomical Observatory, under programs A/2013A/018 and A/2013B/001,and on observations at Cerro Tololo Inter-American Observatory, National Optical Astronomy Observatory (NOAO Prop. IDs 2013A-0618 and 2015A-0618), which is operated by the Association of Universities for Research in Astronomy (AURA) under a cooperative agreement with the National Science Foundation. This project used data obtained with the Dark Energy Camera (DECam), which was constructed by the Dark Energy Survey (DES) collaboration. \\
This paper uses data from observations made with the William Herschel Telescope operated on the island of La Palma by the Isaac Newton Group in the Spanish Observatorio del Roque de los Muchachos of the Instituto de Astrofísica de Canarias. \\
The Millennium Simulation databases used in this paper and the web application providing online access to them were constructed as part of the activities of the German Astrophysical Virtual Observatory (GAVO). \\
O.M. is grateful for the financial support provided by the NEWFELPRO fellowship project in Croatia. \\

\bibliographystyle{mn2e}
\bibliography{biblio}

\subsection*{\rule{\columnwidth}{1pt}}
$^{1}$ Department of Physics and Astronomy, University of Victoria, 3800 Finnerty Road, Victoria, BC, V8P 1A1, Canada \\
$^{2}$ School of Physics and Astronomy, University of Birmingham, Birmingham B152TT, United Kingdom \\
$^{3}$ Australian Astronomical Observatory, PO BOX 915, North Ryde, 1670, Australia \\
$^{4}$ Université Aix-Marseille, CNRS, LAM (Laboratoire d'Astrophysique de Marseille) UMR 7326, 13388, Marseille, France \\
$^{5}$ Argelander-Institut für Astronomie, University of Bonn, Auf dem Hügel 71, 53121 Bonn, Germany \\
$^{6}$ Service d'Astrophysique AIM, CEA Saclay, F-91191 Gif sur Yvette, France \\
$^{7}$ Max-Planck-Institut für extraterrestrische Physik, Giessenbach- straße, D-85748 Garching, Germany  \\
$^{8}$ H.H. Wills Physics Laboratory, University of Bristol, Tyndall Avenue, Bristol, BS8 1TL, UK  \\
$^{9}$ ESA, Villafranca del Castillo, Spain \\
$^{10}$ Department of Astronomy, University of Florida, Gainesville, FL 32611, USA \\
$^{11}$ Astrophysics Research Institute, Liverpool John Moores University, IC2, Liverpool Science Park, 146 Brownlow Hill, Liverpool L3 5RF, United Kingdom \\
$^{12}$ INAF - Observatory of Rome, via Frascati 33, 00040 Monteporzio Catone, Rome, Italy \\
$^{13}$ INAF, IASF Milano, via Bassini 15, I-20133 Milano, Italy \\
$^{14}$ Faculty of Physics, Ludwig Maximilian Universität, München, Germany \\
$^{15}$ Dipartimento di Fisica e Astronomia, Universita` di Bologna, Viale Berti Pichat 6/2, I-40127 Bologna, Italy  \\
$^{16}$ Institut d'Astrophysique Spatiale (IAS), bat. 121, 91405 Orsay Cedex, France \\
$^{17}$ INAF, Osservatorio Astronomico di Brera, Merate, Italy \\
$^{18}$ INAF, Osservatorio Astronomico di Padova, Vicolo dell'Osservatorio, 5, 35122, Padova, Italy \\
$^{19}$ Department of Physics and Astronomy, Macquarie University, Sydney, NSW 2109, Australia \\
$^{20}$ Main Astronomical Observatory, Academy of Sciences of Ukraine, 27 Akademika Zabolotnoho St., 03680 Kyiv, Ukraine \\
$^{21}$ Department of Physics, University of Zagreb, Bijenicka cesta 32, HR-10000 Zagreb, Croatia \\
$^{22}$ Universität Hamburg, Hamburger Sternwarte, Gojenbergsweg 112, 21029 Hamburg, Germany \\
$^{23}$ Max-Planck Institut fuer Kernphysik, Saupfercheckweg 1, 69117 Heidelberg, Germany \\
$^{24}$ Astronomical Observatory, Taras Shevchenko National University of Kyiv,  Observatorna str. 3, 04053 Kyiv, Ukraine \\
$^{25}$ Department of Astronomy, University of Geneva, ch. d’Écogia 16, CH-1290 Versoix \\

\bsp

\label{lastpage}

\appendix
\clearpage
\onecolumn
{
\clearpage
\begin{landscape}
\begin{center}
\section{BCG data}
\LTcapwidth=8in
\renewcommand{\arraystretch}{1.25}
\begin{longtable}{cccccccccccc}
\caption[Summary of XXL-100-GC clusters and BCGs properties]{Summary of XXL-100-GC clusters and BCGs properties. Column 1 shows the clusters unique XXL name; column 2 shows the cluster redshift. Columns 3 and 4 respectively show the mass inside of r$_{500,MT}$ and the value of r$_{500,MT}$ based on \citetalias{xxl_IV} $M-T$ scaling relation. BCG positions are given in columns 5 and 6; column 7 shows the BCG redshift. BCG offset from the X-ray centroid is shown in column 8 and 9. Column 10 shows BCG stellar masses and column 11 gives the \emph{z}-band magnitude difference between the brightest and second brightest cluster members.}
\label{bcg_position}
\\
\hline
   \multicolumn{1}{c}{Name} &
   \multicolumn{1}{c}{z$_{cluster}$} &
   \multicolumn{1}{c}{M$_{500}$} &
   \multicolumn{1}{c}{r$_{500}$} &
   \multicolumn{1}{c}{BCG ra} &
   \multicolumn{1}{c}{BCG dec} &
   \multicolumn{1}{c}{z$_{BCG}$} &
   \multicolumn{2}{c}{BCG offset} &
   \multicolumn{1}{c}{BCG mass} &
   \multicolumn{1}{c}{$\Delta m_{12}$} \\
   
   \multicolumn{1}{c}{} &
   \multicolumn{1}{c}{} &
   \multicolumn{1}{c}{$10^{13}$~M$_{\odot}$} &
   \multicolumn{1}{c}{Mpc} &
   \multicolumn{1}{c}{J2000} &
   \multicolumn{1}{c}{J2000} &
   \multicolumn{1}{c}{} &
   \multicolumn{1}{c}{(")} &
   \multicolumn{1}{c}{r$_{500}$} &
   \multicolumn{1}{c}{$10^{11}$~M$_{\odot}$} &
   \multicolumn{1}{c}{z mag} \\

   \multicolumn{1}{c}{(1)} &
   \multicolumn{1}{c}{(2)} &
   \multicolumn{1}{c}{(3)} &
   \multicolumn{1}{c}{(4)} &
   \multicolumn{1}{c}{(5)} &
   \multicolumn{1}{c}{(6)} &
   \multicolumn{1}{c}{(7)} &
   \multicolumn{1}{c}{(8)} &
   \multicolumn{1}{c}{(9)} &
   \multicolumn{1}{c}{(10)} & 
   \multicolumn{1}{c}{(11)} & \\
   \hline
\endfirsthead

\multicolumn{12}{c}{{\tablename} \thetable{} -- Continued} \\
  \hline
   \multicolumn{1}{c}{Name} &
   \multicolumn{1}{c}{z$_{cluster}$} &
   \multicolumn{1}{c}{M$_{500}$} &
   \multicolumn{1}{c}{r$_{500}$} &
   \multicolumn{1}{c}{BCG ra} &
   \multicolumn{1}{c}{BCG dec} &
   \multicolumn{1}{c}{z$_{BCG}$} &
   \multicolumn{2}{c}{BCG offset} &
   \multicolumn{1}{c}{BCG mass} &
   \multicolumn{1}{c}{$\Delta m_{12}$} \\
   
   \multicolumn{1}{c}{} &
   \multicolumn{1}{c}{} &
   \multicolumn{1}{c}{$10^{13}$~M$_{\odot}$} &
   \multicolumn{1}{c}{Mpc} &
   \multicolumn{1}{c}{J2000} &
   \multicolumn{1}{c}{J2000} &
   \multicolumn{1}{c}{} &
   \multicolumn{1}{c}{(")} &
   \multicolumn{1}{c}{r$_{500}$} &
   \multicolumn{1}{c}{$10^{11}$~M$_{\odot}$} &
   \multicolumn{1}{c}{z mag} \\

   \multicolumn{1}{c}{(1)} &
   \multicolumn{1}{c}{(2)} &
   \multicolumn{1}{c}{(3)} &
   \multicolumn{1}{c}{(4)} &
   \multicolumn{1}{c}{(5)} &
   \multicolumn{1}{c}{(6)} &
   \multicolumn{1}{c}{(7)} &
   \multicolumn{1}{c}{(8)} &
   \multicolumn{1}{c}{(9)} &
   \multicolumn{1}{c}{(10)} & 
   \multicolumn{1}{c}{(11)} & \\
   \hline
\endhead

  \hline
  \multicolumn{12}{l}{{Continued on Next Page\ldots}} \\
\endfoot

  \hline
\endlastfoot


XLSSC 001 	&	0.614	&	 25$\pm12$ 	&	0.777	&	36.2388	&	-3.8147	&	0.617	&	7.6	&	0.067	&	 5.01$^{+0.69}_{-0.51}$ 	&	2.19	\\
XLSSC 003 	&	0.836	&	 19$\pm11$ 	&	0.643	&	36.9092	&	-3.2992	&	0.838	&	1.3	&	0.015	&	 6.42$^{+1.38}_{-1.06}$ 	&	0.07	\\
XLSSC 006 	&	0.429	&	 41$\pm18$ 	&	0.982	&	35.4380	&	-3.7674	&	0.429	&	17.4	&	0.100	&	 12.70$^{+1.22}_{-1.05}$ 	&	0.10	\\
XLSSC 010 	&	0.330	&	 17$\pm 8$ 	&	0.751	&	36.8432	&	-3.3609	&	0.330	&	3.9	&	0.025	&	 6.06$^{+0.54}_{-0.37}$ 	&	1.94	\\
XLSSC 011 	&	0.054	&	 17$\pm 9$ 	&	0.831	&	36.5403	&	-4.9682	&	0.050	&	3.4	&	0.004	&	 2.88$^{+0.22}_{-0.14}$ 	&	0.81	\\
XLSSC 022 	&	0.293	&	 11$\pm 5$ 	&	0.671	&	36.9181	&	-4.8586	&	0.295	&	3.8	&	0.025	&	 6.01$^{+0.53}_{-0.35}$ 	&	0.65	\\
XLSSC 023 	&	0.328	&	 11$\pm 5$ 	&	0.655	&	35.1895	&	-3.4333	&	0.328	&	7.5	&	0.054	&	 6.44$^{+0.58}_{-0.38}$ 	&	0.61	\\
XLSSC 025 	&	0.265	&	 16$\pm 7$ 	&	0.751	&	36.3530	&	-4.6791	&	0.264	&	1.8	&	0.010	&	 6.13$^{+0.54}_{-0.34}$ 	&	1.01	\\
XLSSC 027 	&	0.295	&	 17$\pm 8$ 	&	0.768	&	37.0187	&	-4.8499	&	0.294	&	25.8	&	0.149	&	 5.73$^{+0.51}_{-0.33}$ 	&	0.22	\\
XLSSC 029 	&	1.050	&	 22$\pm12$ 	&	0.626	&	36.0174	&	-4.2240	&	1.050	&	3.8	&	0.050	&	 6.64$^{+2.04}_{-1.49}$ 	&	2.58	\\
XLSSC 036 	&	0.492	&	 24$\pm11$ 	&	0.801	&	35.5286	&	-3.0540	&	0.496	&	5.4	&	0.041	&	 10.30$^{+1.22}_{-0.81}$ 	&	1.05	\\
XLSSC 041 	&	0.142	&	 10$\pm 4$ 	&	0.670	&	36.3782	&	-4.2385	&	0.143	&	1.4	&	0.005	&	 3.51$^{+0.28}_{-0.18}$ 	&	1.41	\\
XLSSC 050 	&	0.141	&	 23$\pm10$ 	&	0.897	&	36.4372	&	-3.2091	&	0.142	&	93.0	&	0.258	&	 5.28$^{+0.41}_{-0.28}$ 	&	0.37	\\
XLSSC 054 	&	0.053	&	 11$\pm 5$ 	&	0.723	&	36.3185	&	-5.8870	&	0.054	&	3.3	&	0.005	&	 3.35$^{+0.27}_{-0.16}$ 	&	0.81	\\
XLSSC 055 	&	0.232	&	 21$\pm10$ 	&	0.843	&	36.4555	&	-5.8962	&	0.233	&	5.9	&	0.026	&	 10.90$^{+0.94}_{-0.60}$ 	&	1.04	\\
XLSSC 056 	&	0.348	&	 22$\pm11$ 	&	0.824	&	33.8676	&	-4.6781	&	0.347	&	18.3	&	0.110	&	 12.20$^{+1.11}_{-0.75}$ 	&	0.94	\\
XLSSC 057 	&	0.153	&	 13$\pm 6$ 	&	0.734	&	34.0505	&	-4.2394	&	0.154	&	8.2	&	0.030	&	 6.49$^{+0.51}_{-0.35}$ 	&	0.63	\\
XLSSC 060 	&	0.139	&	 47$\pm20$ 	&	1.136	&	33.6712	&	-4.5673	&	0.140	&	54.6	&	0.118	&	 14.30$^{+1.07}_{-0.79}$ 	&	0.93	\\
XLSSC 061 	&	0.259	&	 11$\pm 6$ 	&	0.678	&	35.4848	&	-5.7588	&	0.259	&	4.2	&	0.025	&	 8.94$^{+0.79}_{-0.50}$ 	&	2.18	\\
XLSSC 072 	&	1.002	&	 19$\pm11$ 	&	0.613	&	33.8500	&	-3.7256	&	 - 	&	1.9	&	0.025	&	 5.46$^{+1.61}_{-1.08}$ 	&	0.58	\\
XLSSC 083 	&	0.430	&	 37$\pm20$ 	&	0.943	&	32.7350	&	-6.1985	&	0.429	&	4.8	&	0.030	&	 6.40$^{+0.64}_{-0.51}$ 	&	0.01	\\
XLSSC 084 	&	0.430	&	 36$\pm25$ 	&	0.945	&	32.7621	&	-6.2130	&	0.432	&	18.6	&	0.119	&	 3.02$^{+0.30}_{-0.24}$ 	&	0.15	\\
XLSSC 085 	&	0.428	&	 41$\pm27$ 	&	0.976	&	32.8697	&	-6.1963	&	0.429	&	2.9	&	0.018	&	 10.30$^{+1.03}_{-0.81}$ 	&	1.03	\\
XLSSC 087 	&	0.141	&	  8$\pm 3$ 	&	0.619	&	37.7208	&	-4.3478	&	0.141	&	3.4	&	0.014	&	 4.87$^{+0.38}_{-0.26}$ 	&	1.04	\\
XLSSC 090 	&	0.141	&	  4$\pm 2$ 	&	0.507	&	37.1222	&	-4.8565	&	0.142	&	4.4	&	0.022	&	 4.74$^{+0.37}_{-0.25}$ 	&	2.42	\\
XLSSC 091 	&	0.186	&	 51$\pm22$ 	&	1.149	&	37.9215	&	-4.8825	&	0.185	&	17.2	&	0.047	&	 9.32$^{+0.76}_{-0.51}$ 	&	0.49	\\
XLSSC 093 	&	0.429	&	 23$\pm11$ 	&	0.810	&	31.7002	&	-6.9471	&	0.429	&	6.0	&	0.042	&	 6.30$^{+0.62}_{-0.51}$ 	&	0.00	\\
XLSSC 096 	&	0.520	&	 48$\pm31$ 	&	1.000	&	30.9709	&	-5.0279	&	0.521	&	6.9	&	0.043	&	 6.93$^{+0.86}_{-0.56}$ 	&	1.05	\\
XLSSC 097 	&	0.760	&	 32$\pm19$ 	&	0.794	&	33.3426	&	-6.0990	&	0.695	&	4.3	&	0.041	&	 7.48$^{+1.37}_{-1.00}$ 	&	0.06	\\
XLSSC 098 	&	0.297	&	 20$\pm12$ 	&	0.801	&	33.1144	&	-6.0751	&	0.296	&	5.3	&	0.034	&	 7.26$^{+0.64}_{-0.42}$ 	&	1.13	\\
XLSSC 099 	&	0.391	&	 46$\pm40$ 	&	1.032	&	33.2196	&	-6.2033	&	0.361	&	5.7	&	0.029	&	 8.07$^{+0.77}_{-0.57}$ 	&	1.49	\\
XLSSC 100 	&	0.915	&	 26$\pm18$ 	&	0.694	&	31.5473	&	-6.1920	&	0.915	&	6.0	&	0.069	&	 6.27$^{+1.57}_{-1.07}$ 	&	0.62	\\
XLSSC 101 	&	0.756	&	 31$\pm16$ 	&	0.788	&	32.1957	&	-4.4310	&	0.753	&	21.0	&	0.198	&	 13.40$^{+2.40}_{-1.79}$ 	&	1.80	\\
XLSSC 103 	&	0.233	&	 27$\pm17$ 	&	0.913	&	36.8866	&	-5.9644	&	0.232	&	13.7	&	0.056	&	 6.42$^{+0.56}_{-0.35}$ 	&	0.08	\\
XLSSC 104 	&	0.294	&	-	&	1.038	&	37.3287	&	-5.8872	&	0.291	&	31.6	&	0.135	&	 6.75$^{+0.59}_{-0.39}$ 	&	0.12	\\
XLSSC 105 	&	0.429	&	 47$\pm24$ 	&	1.024	&	38.4158	&	-5.5109	&	0.452	&	23.3	&	0.129	&	 4.82$^{+0.48}_{-0.38}$ 	&	0.18	\\
XLSSC 106 	&	0.300	&	 24$\pm11$ 	&	0.856	&	31.3676	&	-5.7324	&	0.302	&	61.1	&	0.320	&	 8.10$^{+0.72}_{-0.47}$ 	&	0.40	\\
XLSSC 107 	&	0.436	&	 16$\pm 8$ 	&	0.711	&	31.3541	&	-7.5945	&	0.439	&	2.1	&	0.017	&	 5.48$^{+0.55}_{-0.45}$ 	&	0.35	\\
XLSSC 108 	&	0.254	&	 13$\pm 6$ 	&	0.705	&	31.8335	&	-4.8252	&	0.255	&	8.9	&	0.051	&	 7.15$^{+0.63}_{-0.40}$ 	&	1.41	\\
XLSSC 109 	&	0.491	&	 23$\pm15$ 	&	0.787	&	32.2967	&	-6.3453	&	0.487	&	3.0	&	0.023	&	 8.37$^{+0.96}_{-0.68}$ 	&	0.32	\\
XLSSC 111 	&	0.299	&	 40$\pm18$ 	&	1.017	&	33.1124	&	-5.6265	&	0.300	&	5.0	&	0.022	&	 11.50$^{+1.05}_{-0.64}$ 	&	0.57	\\
XLSSC 112 	&	0.139	&	  9$\pm 4$ 	&	0.653	&	32.5093	&	-5.4678	&	0.138	&	24.5	&	0.093	&	 5.35$^{+0.41}_{-0.29}$ 	&	0.92	\\
XLSSC 113 	&	0.050	&	  5$\pm 2$ 	&	0.560	&	30.5610	&	-7.0082	&	0.051	&	1.7	&	0.003	&	 3.34$^{+0.26}_{-0.17}$ 	&	0.19	\\
XLSSC 114 	&	0.234	&	 44$\pm51$ 	&	1.070	&	30.4207	&	-5.0302	&	 - 	&	16.8	&	0.059	&	 9.19$^{+0.81}_{-0.50}$ 	&	1.15	\\
XLSSC 115 	&	0.043	&	 12$\pm 7$ 	&	0.740	&	32.6798	&	-6.5797	&	0.043	&	30.3	&	0.035	&	 2.84$^{+0.22}_{-0.14}$ 	&	0.46	\\
XLSSC 502 	&	0.141	&	  5$\pm 2$ 	&	0.532	&	348.4413	&	-53.4368	&	0.140	&	5.0	&	0.023	&	 6.29$^{+0.48}_{-0.33}$ 	&	1.47	\\
XLSSC 503 	&	0.336	&	 10$\pm 5$ 	&	0.642	&	350.6469	&	-52.7470	&	0.334	&	3.8	&	0.029	&	 5.70$^{+0.51}_{-0.34}$ 	&	0.26	\\
XLSSC 505 	&	0.055	&	  9$\pm 4$ 	&	0.661	&	352.2513	&	-52.2364	&	0.055	&	6.7	&	0.011	&	 8.27$^{+0.65}_{-0.40}$ 	&	0.76	\\
XLSSC 507 	&	0.566	&	 12$\pm 6$ 	&	0.612	&	353.3732	&	-52.2537	&	0.569	&	7.4	&	0.080	&	 8.18$^{+1.07}_{-0.67}$ 	&	0.12	\\
XLSSC 509 	&	0.633	&	 29$\pm17$ 	&	0.806	&	356.4538	&	-54.0466	&	0.635	&	26.7	&	0.230	&	 4.50$^{+0.61}_{-0.49}$ 	&	0.02	\\
XLSSC 510 	&	0.395	&	 15$\pm 7$ 	&	0.711	&	357.5395	&	-55.3331	&	0.395	&	2.3	&	0.018	&	 5.29$^{+0.50}_{-0.37}$ 	&	1.59	\\
XLSSC 511 	&	0.130	&	  5$\pm 2$ 	&	0.545	&	357.7522	&	-55.3704	&	0.133	&	3.7	&	0.016	&	 2.77$^{+0.21}_{-0.14}$ 	&	0.38	\\
XLSSC 512 	&	0.402	&	 26$\pm12$ 	&	0.848	&	352.4831	&	-56.1357	&	0.402	&	2.3	&	0.014	&	 9.51$^{+0.90}_{-0.69}$ 	&	1.00	\\
XLSSC 513 	&	0.378	&	 34$\pm17$ 	&	0.936	&	349.2161	&	-54.8990	&	0.377	&	21.7	&	0.121	&	 11.10$^{+1.05}_{-0.70}$ 	&	0.45	\\
XLSSC 514 	&	0.169	&	  7$\pm 3$ 	&	0.582	&	351.3990	&	-54.7208	&	0.169	&	12.0	&	0.060	&	 4.30$^{+0.34}_{-0.23}$ 	&	0.55	\\
XLSSC 515 	&	0.101	&	  5$\pm 2$ 	&	0.540	&	351.4173	&	-54.7419	&	0.100	&	6.7	&	0.023	&	 5.62$^{+0.43}_{-0.28}$ 	&	1.03	\\
XLSSC 517 	&	0.699	&	 20$\pm12$ 	&	0.698	&	350.4494	&	-55.9704	&	0.697	&	1.1	&	0.012	&	 6.34$^{+0.90}_{-0.83}$ 	&	0.12	\\
XLSSC 518 	&	0.177	&	  5$\pm 2$ 	&	0.535	&	349.8214	&	-55.3243	&	0.177	&	3.9	&	0.022	&	 7.71$^{+0.61}_{-0.42}$ 	&	0.87	\\
XLSSC 519 	&	0.270	&	  6$\pm 3$ 	&	0.555	&	353.0194	&	-55.2123	&	0.270	&	2.3	&	0.017	&	 4.99$^{+0.43}_{-0.28}$ 	&	0.92	\\
XLSSC 520 	&	0.175	&	 17$\pm 7$ 	&	0.805	&	352.5017	&	-54.6188	&	0.176	&	0.8	&	0.003	&	 11.00$^{+0.82}_{-0.64}$ 	&	2.10	\\
XLSSC 521 	&	0.807	&	 31$\pm18$ 	&	0.775	&	352.1791	&	-55.5669	&	0.807	&	0.4	&	0.004	&	 13.70$^{+2.74}_{-2.13}$ 	&	0.84	\\
XLSSC 522 	&	0.395	&	 15$\pm 7$ 	&	0.711	&	351.6401	&	-55.0199	&	0.395	&	10.3	&	0.078	&	 5.62$^{+0.53}_{-0.40}$ 	&	0.28	\\
XLSSC 523 	&	0.343	&	 19$\pm10$ 	&	0.779	&	350.5019	&	-54.7499	&	0.345	&	3.0	&	0.019	&	 4.76$^{+0.42}_{-0.29}$ 	&	0.21	\\
XLSSC 524 	&	0.270	&	 16$\pm 8$ 	&	0.754	&	353.0646	&	-54.7032	&	0.269	&	11.4	&	0.063	&	 6.30$^{+0.54}_{-0.35}$ 	&	0.28	\\
XLSSC 525 	&	0.379	&	 24$\pm10$ 	&	0.832	&	349.3403	&	-53.9612	&	0.371	&	6.9	&	0.044	&	 9.22$^{+0.85}_{-0.61}$ 	&	0.45	\\
XLSSC 527 	&	0.076	&	 24$\pm27$ 	&	0.926	&	349.5734	&	-55.9839	&	0.076	&	13.3	&	0.021	&	 6.47$^{+0.51}_{-0.31}$ 	&	0.41	\\
XLSSC 528 	&	0.302	&	 23$\pm12$ 	&	0.839	&	349.6818	&	-56.2034	&	0.303	&	3.2	&	0.017	&	 9.91$^{+0.86}_{-0.57}$ 	&	0.69	\\
XLSSC 529 	&	0.547	&	 23$\pm11$ 	&	0.769	&	349.7037	&	-56.2865	&	0.548	&	15.6	&	0.131	&	 6.38$^{+0.82}_{-0.51}$ 	&	0.62	\\
XLSSC 530 	&	0.182	&	 11$\pm 5$ 	&	0.686	&	348.8342	&	-54.3440	&	0.190	&	5.7	&	0.026	&	 6.04$^{+0.48}_{-0.33}$ 	&	0.29	\\
XLSSC 531 	&	0.391	&	 38$\pm30$ 	&	0.966	&	349.8752	&	-56.6495	&	0.390	&	3.6	&	0.020	&	 10.00$^{+0.98}_{-0.66}$ 	&	0.24	\\
XLSSC 532 	&	0.392	&	 19$\pm10$ 	&	0.772	&	352.9477	&	-52.6657	&	0.391	&	11.0	&	0.077	&	 5.69$^{+0.54}_{-0.40}$ 	&	0.32	\\
XLSSC 533 	&	0.107	&	 15$\pm 6$ 	&	0.789	&	351.7243	&	-52.6971	&	0.108	&	46.0	&	0.115	&	 4.60$^{+0.35}_{-0.23}$ 	&	0.04	\\
XLSSC 534 	&	0.853	&	 27$\pm18$ 	&	0.725	&	350.1089	&	-53.3587	&	0.853	&	12.5	&	0.131	&	 6.57$^{+1.44}_{-1.11}$ 	&	0.43	\\
XLSSC 535 	&	0.172	&	 14$\pm 6$ 	&	0.756	&	351.5538	&	-53.3162	&	0.171	&	1.7	&	0.006	&	 9.30$^{+0.73}_{-0.51}$ 	&	0.51	\\
XLSSC 537 	&	0.515	&	 39$\pm21$ 	&	0.934	&	354.0297	&	-53.8766	&	0.517	&	3.4	&	0.023	&	 13.00$^{+1.56}_{-1.03}$ 	&	1.86	\\
XLSSC 538 	&	0.332	&	 20$\pm12$ 	&	0.804	&	354.6477	&	-54.6242	&	0.332	&	6.8	&	0.041	&	 9.88$^{+0.87}_{-0.59}$ 	&	0.41	\\
XLSSC 539 	&	0.184	&	  5$\pm 2$ 	&	0.520	&	355.7959	&	-55.8814	&	0.182	&	5.4	&	0.030	&	 7.26$^{+0.59}_{-0.39}$ 	&	0.78	\\
XLSSC 540 	&	0.414	&	 20$\pm 9$ 	&	0.776	&	355.6308	&	-56.3532	&	0.411	&	4.9	&	0.035	&	 9.01$^{+0.86}_{-0.68}$ 	&	0.86	\\
XLSSC 541 	&	0.188	&	 18$\pm 8$ 	&	0.805	&	355.4330	&	-55.9637	&	0.188	&	8.2	&	0.032	&	 5.54$^{+0.45}_{-0.29}$ 	&	0.51	\\
XLSSC 542 	&	0.402	&	 74$\pm32$ 	&	1.202	&	353.1145	&	-53.9744	&	0.405	&	8.7	&	0.039	&	 14.40$^{+1.37}_{-1.04}$ 	&	0.97	\\
XLSSC 543 	&	0.381	&	 14$\pm 7$ 	&	0.689	&	354.8637	&	-55.8407	&	0.383	&	10.1	&	0.077	&	 5.70$^{+0.53}_{-0.38}$ 	&	0.33	\\
XLSSC 544 	&	0.095	&	 15$\pm 7$ 	&	0.788	&	349.8155	&	-53.5330	&	0.096	&	2.7	&	0.006	&	 8.17$^{+0.63}_{-0.40}$ 	&	0.17	\\
XLSSC 546 	&	0.792	&	 20$\pm10$ 	&	0.668	&	352.4201	&	-53.2489	&	0.860	&	13.6	&	0.154	&	 4.51$^{+0.87}_{-0.67}$ 	&	0.30	\\
XLSSC 547 	&	0.371	&	 32$\pm18$ 	&	0.920	&	351.4277	&	-53.2768	&	0.370	&	2.1	&	0.010	&	 11.60$^{+1.05}_{-0.74}$ 	&	0.98	\\
XLSSC 550 	&	0.109	&	  3$\pm 2$ 	&	0.480	&	352.2079	&	-52.5770	&	0.107	&	8.5	&	0.035	&	 5.34$^{+0.41}_{-0.27}$ 	&	1.42	\\

\end{longtable}
\end{center}
\end{landscape}
}

\end{document}